\documentclass[
  aps,
  prx,
  reprint,
  superscriptaddress,
  longbibliography,
  floatfix
]{revtex4-2}

\usepackage{amssymb,amsthm,mathtools}
\usepackage{microtype}
\usepackage{booktabs,array,tabularx,adjustbox}
\usepackage{graphicx}
\usepackage{xcolor}
\usepackage{placeins}
\usepackage[hidelinks]{hyperref}
\usepackage[nameinlink,capitalise,noabbrev]{cleveref}

\usepackage{amsmath,amsfonts,bm}

\def\eqref#1{equation~\ref{#1}}

\def\1{\bm{1}}

\DeclareMathAlphabet{\mathsfit}{\encodingdefault}{\sfdefault}{m}{sl}
\SetMathAlphabet{\mathsfit}{bold}{\encodingdefault}{\sfdefault}{bx}{n}

\DeclareMathOperator{\Tr}{Tr}

\allowdisplaybreaks

\newcommand{\cA}{\mathcal A}

\newcommand{\cC}{\mathcal C}

\newcommand{\cE}{\mathcal E}
\newcommand{\cF}{\mathcal F}
\newcommand{\cH}{\mathcal H}
\newcommand{\cK}{\mathcal K}

\newcommand{\cN}{\mathcal N}
\newcommand{\cR}{\mathcal R}
\newcommand{\cS}{\mathcal S}

\newcommand{\cV}{\mathcal V}

\newcommand{\bbE}{\mathbb E}
\newcommand{\bbP}{\mathbb P}
\newcommand{\TV}{\operatorname{TV}}
\newcommand{\Sh}{\operatorname{Sh}}
\newcommand{\LOCC}{\operatorname{LOCC}}

\newcommand{\id}{\operatorname{id}}
\newcommand{\diag}{\operatorname{diag}}

\newcommand{\norm}[1]{\left\lVert #1\right\rVert}
\newcommand{\abs}[1]{\left\lvert #1\right\rvert}

\newcommand{\ketbra}[2]{\left|#1\right\rangle\!\left\langle#2\right|}
\newcommand{\ket}[1]{\left|#1\right\rangle}
\newcommand{\bra}[1]{\left\langle#1\right|}
\newcommand{\raqdv}{\textsc{RA-QDV}}

\newtheorem{theorem}{Theorem}[section]
\newtheorem{proposition}[theorem]{Proposition}
\newtheorem{corollary}[theorem]{Corollary}
\newtheorem{lemma}[theorem]{Lemma}
\newtheorem{assumption}[theorem]{Assumption}
\crefname{assumption}{Assumption}{Assumptions}
\Crefname{assumption}{Assumption}{Assumptions}
\theoremstyle{definition}
\newtheorem{definition}[theorem]{Definition}
\crefname{definition}{Definition}{Definitions}
\Crefname{definition}{Definition}{Definitions}
\theoremstyle{remark}
\newtheorem{remark}[theorem]{Remark}

\begin{document}

\title{Shapley Valuation of Finite-Copy Quantum Data Depends on Physical Access}

\author{Qipeng Qian}
\email{qianqipeng@supcon.com}
\affiliation{SUPCON Technology, Hangzhou, China}
\affiliation{College of Artificial Intelligence, Zhejiang University, Hangzhou 310027, China}

\author{Yuntao Qian}
\email{ytqian@zju.edu.cn}
\affiliation{College of Artificial Intelligence, Zhejiang University, Hangzhou 310027, China}


\begin{abstract}
Data valuation asks how learning utility should be attributed to training data contributors. Most classical formulations begin after data have become reusable records, so the physical readout of the data is effectively fixed. Finite-copy quantum data are different: unknown states are consumable physical systems, and the same supplied states and downstream task can yield different Shapley values under different physical access models. Our framework makes this dependence explicit by treating physical access as a component of quantum data valuation itself. 
We establish an exact connection between physical-access advantage and contributor-level data valuation. For nested access models, we prove that the maximal downstream utility gain enabled by richer physical access exactly determines the largest symmetric Shapley ranking-reversal margin. More generally, for arbitrary access-model pairs, including non-nested ones, we derive an exact geometric characterization of the possible shifts of the full Shapley attribution vector. For fixed learning pipelines, we further obtain an operational Shapley-observable representation for finite-copy valuation. 
Numerical experiments demonstrate that identical quantum samples can receive different values and rankings when only the physical access model is changed. These results establish that quantum data value is not an intrinsic property of the underlying states alone, but emerges from the interaction between quantum states, physical access, and the downstream learning task. 
\end{abstract}

\maketitle

\section{Introduction}
\label{sec:introduction}

Data valuation studies how the utility of a learned model should be attributed to individual training data contributors \citep{sim2022datavaluation}. This perspective is important for data selection, acquisition, quality assessment, and incentive mechanisms, where different samples may contribute unequally to the final model \citep{koh2017influence,ai2025instrumental}. Data Shapley formalizes this idea through average marginal contributions \citep{shapley1953value,ghorbani2019datashapley,jia2019efficient}. 
Subsequent work has developed more scalable or robust valuation schemes, including distributional and alternative semivalue formulations \citep{ghorbani2020distributional,kwon2022beta,wang2023banzhaf}, as well as efficient estimators based on reinforcement learning, out-of-bag evaluation, or a single training run \citep{yoon2020dvrl,kwon2023dataoob,wang2025inrun}. Other work has examined the dependence of data values on the downstream learner \citep{just2023lava} and on the specification of the utility itself \citep{wang2024rethinking,tamine2026utility,diehl2025arbitrary}. 
However, most such formulations begin after the data have already become reusable classical records. The representation of each sample and the way in which the learner can read it are therefore treated as fixed, rather than as part of the valuation problem itself.

Finite-copy quantum data do not admit this simplification. Unknown quantum states are consumable physical systems, and task-relevant information must first be extracted through an admissible measurement procedure before it becomes classically available to the learner. Quantum learning with limited copies has therefore emphasized learning performance, generalization, learnability, and resource requirements \citep{banchi2023statistical,caro2024generalization,gilboa2025consumable,noller2025hierarchy,chen2026fewcopy}. 
A direct consequence is that two learners supplied with the same physical states need not have access to the same effective information: different measurement models can expose genuinely different classical information to the same downstream decision problem \citep{banchi2023statistical,caro2024generalization,noller2025hierarchy,takagi2019resource,bennett1999nonlocality,walgate2000local}. 
Thus, for finite-copy quantum data, the value of a sample depends not only on the state supplied, but also on the physical access model used to extract task-relevant information from it. 

This observation connects data valuation to quantum statistical comparison and resource theories. Statistical comparison and randomization frameworks characterize when one quantum experiment can simulate another \citep{blackwell1953,jencova2016comparison,jencova2020general}, while resource-theoretic approaches quantify operational advantages associated with measurements or other quantum resources \citep{chitambar2019resource,takagi2019resource,skrzypczyk2019incompatible}. These frameworks can answer whether one physical access model provides more useful information or better decision performance than another. What they do not determine is how that access advantage should be distributed among the individual data contributors whose data jointly produce the learning utility. Related ideas appear in resource-dependent cooperative quantum games, where the available strategy or correlation resource can change the value of a game \citep{crann2023cooperative}, while prior quantum Shapley work has focused on acceleration or explainability rather than access-dependent data valuation \citep{burge2024quantumshapley}. 

The unresolved issue is therefore data-contributor-level attribution. Suppose the supplied quantum states and the downstream task are held fixed while only the physical access model is changed. Existing comparisons of access models can quantify how much the best achievable task utility changes, but this scalar difference does not determine how the value of the training data should be distributed among individual data contributors. In particular, it does not reveal which data contributors become more or less valuable, or whether their relative ordering changes. Such data-contributor-level information is essential when data valuation is used to guide data selection, acquisition, quality assessment, or resource allocation. To our knowledge, existing quantum-data-valuation frameworks do not treat physical access itself as part of the valuation problem, nor characterize how changes in physical access reshape the values assigned to individual data contributors. 

In this work, we develop an access-aware Shapley framework for finite-copy quantum data. We keep the supplied quantum states and the downstream task fixed and vary only the physical access model. This isolates a specifically quantum source of valuation dependence: the same data can induce different coalition values because different admissible measurements reveal different information. 
The resulting quantum Shapley value is therefore not an intrinsic property of a quantum state alone, but a property of the interaction among the supplied quantum states, physical access, and the downstream task. 

Our contributions are summarized as follows:
\begin{itemize}
    \item We formulate an access-aware Shapley valuation framework for finite-copy quantum data, treating the physical access model as an explicit component of the valuation problem.

    \item We establish exact theoretical relations between physical-access advantage and contributor-level valuation. For nested access models, we characterize the strongest access-induced Shapley ranking reversal; for arbitrary access-model pairs, we derive a geometric characterization of the possible shifts of the full Shapley attribution vector. 

    \item For fixed learning pipelines, we derive a Shapley-observable representation that provides an operational route to finite-copy valuation. We validate the framework numerically on random quantum discrimination tasks, a controlled TFIM learning pipeline, and a quantum convolutional neural network (QCNN), where changing only the allowed physical access changes data-contributor values and rankings.
\end{itemize}

The remainder of the paper is organized as follows.
Section~\ref{sec:setup} introduces the data-contributor setting, downstream task, physical access models, and access-aware Shapley valuation. 
Section~\ref{sec:normative} develops the main theoretical characterization of how physical access changes Shapley value, including ranking reversals and the geometry of full attribution-vector shifts. 
Section~\ref{sec:operational} turns to fixed learning pipelines and derives the operational Shapley-observable representation for finite-copy valuation. 
Section~\ref{sec:experiments} presents the numerical experiments, including random quantum discrimination, the controlled TFIM benchmark, and the QCNN study. 
Finally, Section~\ref{sec:discussion} summarizes the implications of access-aware quantum data valuation and outlines directions for future theoretical and practical development.

\section{What Are We Valuing?}
\label{sec:setup}

The question in this paper is simple to state. Several contributors supply quantum training data for the same downstream task. We assign each contributor a Shapley value. We then change only the \emph{physical way in which the quantum data may be read} and ask: does the attribution change, and by how much?

\paragraph{Contributors and quantum data.}
Let $N=[n]=\{1,\ldots,n\}$ index the contributors, and let $S\subseteq N$ denote any subset of contributors. In standard Shapley-value terminology, such a subset is called a \emph{coalition}. 

The contributors determine which quantum data are available, while $\theta\in\Theta$ represents the unknown quantity relevant to the downstream task. When the true value is $\theta$, we denote the quantum state available from contributor subset $S$ by
\[
\rho_{\theta,S}.
\]
Thus, the index $S$ specifies \emph{which contributors provide the data}, whereas $\theta$ specifies \emph{which underlying instance of the task generated the quantum state}.

For each subset $S$, we collect all states that may arise as the unknown parameter $\theta$ varies into
\begin{equation}
    \mathcal E_S
    :=
    \left\{
        \rho_{\theta,S}
    \right\}_{\theta\in\Theta}.
    \label{eq:coalition-experiment}
\end{equation}
We call $\mathcal E_S$ the \emph{quantum data family} associated with $S$ (formally, a quantum statistical experiment). It describes the quantum data that may be available when exactly the contributors in $S$ participate, before specifying how those data are processed or what value they provide for the downstream task.

A common physical realization is that contributor $i$ supplies a quantum register with Hilbert space $\mathcal H_i$. If all contributors are present, the joint state lives on
\[
\mathcal H_N=\bigotimes_{i\in N}\mathcal H_i.
\]
When only contributors in $S$ are retained, the unavailable registers are discarded, giving the reduced state
\[
\rho_{\theta,S}
=
\Tr_{S^c}\!\left[\rho_{\theta,N}\right].
\]
This subsystem picture is not required for the general results below, but it is the concrete setting used again in \cref{sec:operational} when we study one fixed finite-copy training block.

\paragraph{The downstream task.}
A task specifies what the learner is trying to infer or decide. Formally, $\theta$ has prior $\pi$, the learner outputs an action $a$ from a finite alphabet $\mathsf A$, and receives a payoff $u(\theta,a)\in[0,1]$ (equivalently, loss $\ell=1-u\in[0,1]$). We write $t=(\pi,u)$ when the task itself needs to be explicit. Nothing in the comparison below changes the supplied states or this task.

\paragraph{The only variable we change: physical access.}
A \emph{physical access model} $\cR$ specifies which quantum processing and measurement protocols the learner is allowed to use. We simply call it an \emph{access model} below. Examples include one-way LOCC \citep{chitambar2014locc}, a bounded-copy measurement family, or unrestricted global measurements. Thus two access models may receive exactly the same state $\rho_{\theta,S}$ but expose different information to the learner. If $\cR_0\subseteq\cR_1$, every protocol allowed under $\cR_0$ is also allowed under $\cR_1$.

\begin{assumption}[Shared baseline and free classical processing]
\label{ass:shared-baseline}
Compared access models differ only in how they process quantum training systems. They share the same optimal no-data risk $R_{\emptyset}$. They may freely discard supplied systems, and may read and post-process explicitly classical registers without restriction.
\end{assumption}

\paragraph{Value of a contributor subset.}
We first study the best value that a contributor subset \emph{could} obtain under a physical access model. Let $R_{\cR}^{\star}(S)$ be the minimum expected test risk achievable from subset $S$ using a protocol allowed by $\cR$. The value of subset $S$ is the improvement over having no training data:
\begin{definition}[Best-achievable subset value]
\label{def:normative-value}
\begin{align}
R_{\cR}^{\star}(S)&=\inf_{\cA\in\cR}R_{\mathrm{test}}(\cA;S),\\
v_{\cR}^{\star}(S)&=R_{\emptyset}-R_{\cR}^{\star}(S),
\quad
v_{\cR}^{\star}(\varnothing)=0.
\label{eq:normative-value}
\end{align}
\end{definition}
The star $\star$ indicates that we optimize over all learners allowed by the access model. Thus, for each subset we ask for the best performance the access model makes possible, rather than committing to one deployed learner.

\paragraph{Contributor value.}
Shapley value turns these subset values into one number per contributor. Contributor $i$ receives its average marginal improvement when it is added to the contributors that appear before it in a random ordering:
\begin{align}
\phi_i^{\cR,\star}
&=\Sh_i(v_{\cR}^{\star})
:=\sum_{S\subseteq N\setminus\{i\}}w_S
\big[v_{\cR}^{\star}(S\cup\{i\})-v_{\cR}^{\star}(S)\big],\\
w_S&=\frac{|S|!(n-|S|-1)!}{n!}.\nonumber
\label{eq:normative-shapley}
\end{align}
Equivalently, put the contributors in a uniformly random order and measure how much contributor $i$ improves the subset that appears before it; $\phi_i^{\cR,\star}$ is the average of that improvement. We use the same operator $\Sh_i(g)$ for any set function $g$.

At this point the object of interest is already clear: changing $\cR$ changes subset values, which can change the Shapley vector $\boldsymbol\phi^{\cR,\star}$. The next section asks how large that change can be and how it is distributed across contributors. A second notion---the value realized by one \emph{fixed} learner---is introduced only later in \cref{sec:operational}, where it is actually needed.

\section{How Physical Access Changes Shapley Value}
\label{sec:normative}

There are two levels to the comparison. First, we use a scalar warm-up: can one access model extract more task-relevant information than another, and can that advantage reverse a Shapley ranking? Second, we characterize the \emph{entire vector} of possible contributor-value changes. The second statement is the main result.

\subsection{A scalar warm-up: access advantage and ranking reversal}

Before studying how an access-model change redistributes value across many contributors, it is useful to answer a simpler question: if two learners receive the same quantum data but are allowed different physical measurements, how much more useful can the richer access be? This scalar comparison will later serve as the calibration for a Shapley ranking reversal.

We first describe what an access model exposes to the downstream classical decision problem. Let
\[
\cE=\{\rho_\theta\}_{\theta\in\Theta}
\]
be a quantum data family, let $\cR$ be an access model, i.e., a class of allowed quantum measurement protocols, and let $\mathsf A$ be the finite set of possible classical outputs or decisions. We write $\Delta(\mathsf A)$ for the set of probability distributions over $\mathsf A$.

An allowed measurement protocol $M\in\cR$ produces an output $a\in\mathsf A$. If the underlying state is $\rho_\theta$, the resulting conditional probability is denoted by $P^M(a\mid\theta)$. For an effective positive-operator-valued measure (POVM) $M=\{M_a\}_{a\in\mathsf A}$, the Born rule gives 

\begin{equation}
P^M(a\mid\theta)=\Tr(M_a\rho_\theta).
\label{eq:measurement-induced-distribution}
\end{equation}
We use $P^M$ for the entire conditional behavior, meaning the collection of output distributions $\{P^M(\cdot\mid\theta)\}_{\theta\in\Theta}$.

The set of all classical behaviors obtainable from $\cE$ under access model $\cR$ is
\begin{equation}
\mathfrak P_{\cR}^{\mathsf A}(\cE)
=
\overline{\operatorname{conv}}
\{P^M:M\in\cR\}
\subseteq
\prod_{\theta\in\Theta}\Delta(\mathsf A).
\label{eq:interface-set}
\end{equation}
We call this the \emph{accessible-output set}, because it contains every classical output behavior that the access model can make available from the same quantum data. The product on the right simply means that one element of the accessible-output set specifies one output distribution for every possible value of $\theta$. The convex hull allows free classical randomization between allowed measurement protocols, and the closure includes limiting behaviors. Operationally, $\mathfrak P_{\cR}^{\mathsf A}(\cE)$ is everything the downstream classical decision problem can see after the quantum data have been processed using access model $\cR$.

Now suppose
\[
\cR_0\subseteq\cR_1,
\]
so every protocol allowed under $\cR_0$ is also allowed under $\cR_1$, while $\cR_1$ may allow additional operations. Then
\[
\mathfrak P_{\cR_0}^{\mathsf A}(\cE)
\subseteq
\mathfrak P_{\cR_1}^{\mathsf A}(\cE).
\]
The richer access model may therefore expose classical behaviors that the weaker access model cannot reproduce exactly.

To quantify this, take any behavior $P\in\mathfrak P_{\cR_1}^{\mathsf A}(\cE)$ produced by the richer access model and ask how closely it can be simulated by some $Q\in\mathfrak P_{\cR_0}^{\mathsf A}(\cE)$. For a fixed $\theta$, their output distributions are compared by total variation,
\begin{equation}
\TV\!\left(P(\cdot\mid\theta),Q(\cdot\mid\theta)\right)
=
\frac12\sum_{a\in\mathsf A}
\left|P(a\mid\theta)-Q(a\mid\theta)\right|.
\label{eq:tv-conditional}
\end{equation}
The simulation should work for every possible underlying value $\theta$, so we use the worst-case distance over $\Theta$. We then let the weaker access model choose its best simulation and, finally, choose the richer-access behavior that is hardest to simulate. This gives
\begin{equation}
\delta_{0\to1}^{\mathsf A}(\cE)
=
\sup_{P\in\mathfrak P_{\cR_1}^{\mathsf A}(\cE)}
\inf_{Q\in\mathfrak P_{\cR_0}^{\mathsf A}(\cE)}
\max_{\theta\in\Theta}
\TV\!\left(P(\cdot\mid\theta),Q(\cdot\mid\theta)\right).
\label{eq:resource-deficiency}
\end{equation}
Mathematically, this is the standard one-sided deficiency. Throughout the paper, we call it the \emph{resource simulation gap}, because it measures how well the weaker access model can simulate everything exposed by the richer one. If $\delta_{0\to1}^{\mathsf A}(\cE)=0$, every classical behavior obtainable with $\cR_1$ can also be reproduced by $\cR_0$; a positive value means that the richer physical access exposes genuinely new classical information.

The same gap can be viewed through an actual downstream task. Recall that a task is $t=(\pi,u)$, where $\pi_\theta$ is the prior probability of $\theta$ and $u(\theta,a)\in[0,1]$ is the payoff for taking action $a$ when the true value is $\theta$. Under access model $\cR$, the best expected payoff obtainable from $\cE$ is
\begin{equation}
U_{\cR}(\pi,u;\cE)
=
\sup_{P\in\mathfrak P_{\cR}^{\mathsf A}(\cE)}
\sum_{\theta\in\Theta}\sum_{a\in\mathsf A}
\pi_\theta u(\theta,a)P(a\mid\theta).
\label{eq:resource-task-utility}
\end{equation}
Thus
\[
U_{\cR_1}(\pi,u;\cE)-U_{\cR_0}(\pi,u;\cE)
\]
is the additional task performance made possible by the richer access model for this particular task. Maximizing this difference over all normalized priors and payoffs asks for the task that benefits most from the additional access.

We now translate this resource simulation gap into Shapley language. Regard the quantum data family $\cE$ as the data supplied by one access-sensitive contributor $B$. Introduce a second contributor $C$ that supplies a finite classical register which both access models can read equally well. We also take $C$ to be conditionally independent of $B$ given $\theta$, so that $C$ acts only as an access-independent reference rather than changing the physical access available to $B$. Its singleton value is therefore the same under $\cR_0$ and $\cR_1$.

The purpose of $C$ is only to provide a ruler. A strict ranking reversal occurs when
\begin{equation}
\phi_C^{\cR_0,\star}>\phi_B^{\cR_0,\star}
\quad\text{but}\quad
\phi_B^{\cR_1,\star}>\phi_C^{\cR_1,\star}.
\label{eq:benchmark-reversal-condition}
\end{equation}
The smaller of these two positive ranking gaps is the symmetric reversal margin. The next theorem says that this margin is calibrated exactly by the resource simulation gap above.

\begin{theorem}[Access advantage and Shapley ranking reversal]
\label{thm:valuation-deficiency-duality}
For every finite $\cE$, finite $\mathsf A$, and nested access models satisfying Assumption~\ref{ass:shared-baseline},
\begin{equation}
\delta_{0\to1}^{\mathsf A}(\cE)
=
\sup_{\substack{\pi\in\Delta(\Theta)\\0\le u\le1}}
\left[
U_{\cR_1}(\pi,u;\cE)-U_{\cR_0}(\pi,u;\cE)
\right].
\label{eq:valuation-deficiency-duality}
\end{equation}
Moreover, for every fixed normalized task, there exists an access-independent finite classical reference contributor $C$ for which
\begin{align}
\phi_C^{\cR_0,\star}-\phi_B^{\cR_0,\star}
&=
\phi_B^{\cR_1,\star}-\phi_C^{\cR_1,\star}\nonumber\\
&=
\frac12\left[
U_{\cR_1}(\pi,u;\cE)-U_{\cR_0}(\pi,u;\cE)
\right].
\label{eq:taskwise-reversal-calibration}
\end{align}
No access-independent classical reference can make the smaller of the two reversal gaps larger than the right-hand side of \cref{eq:taskwise-reversal-calibration}.
\end{theorem}

The first statement is the standard deficiency--decision duality: the resource simulation gap is exactly the largest performance improvement that the richer access model can provide over all downstream tasks.

The second statement connects this access advantage to Shapley attribution. Consider any fixed task for which $\cR_1$ performs better than $\cR_0$. Then one can choose an access-independent classical reference contributor such that the access-sensitive contributor is ranked below the reference under $\cR_0$ but above it under $\cR_1$. In other words, an improvement that comes purely from richer physical access can appear directly as a Shapley ranking reversal.

The theorem also quantifies the strength of this reversal. The reference can be chosen so that the two ranking gaps are equal, with each gap equal to one half of the task-specific access advantage. Consequently, whenever $\delta_{0\to1}^{\mathsf A}(\cE)>0$, there exists at least one downstream task for which a strict Shapley ranking reversal can occur. The construction of the reference and the proof are given in \cref{app:duality-corollaries,app:proof-valuation-deficiency}.

\subsection{Vector extension to multiple contributors}
\label{sec:shapley-interface-geometry}

The previous subsection showed that changing the physical access model can reverse a contributor ranking. That result focuses on one access-sensitive contributor relative to one reference. With many contributors, however, an access change can affect several Shapley values at once. We therefore extend the same idea from a single ranking comparison to the whole Shapley vector.

For a fixed downstream task $t$, define the \emph{Shapley shift vector}
\begin{equation}
\boldsymbol\Delta_{\cR\to\cR'}(t)
:=
\boldsymbol\phi^{\cR',\star}(t)-\boldsymbol\phi^{\cR,\star}(t).
\label{eq:valuation-shift-vector}
\end{equation}
Its $i$th coordinate is simply the change in contributor $i$'s Shapley value when the access model changes from $\cR$ to $\cR'$. As the downstream task varies, the same pair of access models can produce different shift vectors. We collect these possibilities into the \emph{Shapley shift set}
\begin{equation}
\cV_{\cR\to\cR'}^{\mathsf A}
:=
\overline{\operatorname{conv}}
\left\{
\boldsymbol\Delta_{\cR\to\cR'}(t):t\text{ normalized}
\right\}.
\label{eq:valuation-shift-body}
\end{equation}
The convex hull is used only to package all extremal linear questions about the shift vector into one closed set.

A direction $\xi\in\mathbb R^n$ specifies which such question we ask. Here $e_i$ denotes the $i$th standard basis vector. Taking $\xi=e_i$ asks how much contributor $i$ can gain; $\xi=-e_i$ asks how much it can lose; and $\xi=e_i-e_j$ asks how much the ranking gap between contributors $i$ and $j$ can move. Thus the direction $\xi$ introduces no new physical assumption---it simply selects one linear feature of the same Shapley shift vector.

\begin{theorem}[Vector extension of access-induced Shapley shifts]
\label{thm:valuation-shift-body}
For arbitrary admissible physical access models $\cR$ and $\cR'$, including pairs for which neither contains the other, the largest possible directional Shapley change
\begin{equation}
\sup_{t\text{ normalized}}
\xi^\top\boldsymbol\Delta_{\cR\to\cR'}(t)
\label{eq:max-directional-shift-main}
\end{equation}
is determined exactly by the accessible-output sets
$\mathfrak P_{\cR}^{\mathsf A}(\cE_S)$ and
$\mathfrak P_{\cR'}^{\mathsf A}(\cE_S)$ across contributor subsets, combined with the same Shapley weights that define the original attribution. Equivalently, it is an exact directed simulation gap between the positive and negative Shapley-weighted contributions. The explicit construction and equality are given in \cref{eq:directional-shift-geometry}.
\end{theorem}

The theorem is the vector version of the ranking-reversal result above. Instead of asking only whether one contributor can cross one reference, we may ask how far any chosen linear feature of the full Shapley vector can move. The choices $\xi=e_i$, $-e_i$, and $e_i-e_j$ give, respectively, the largest possible increase of one contributor, decrease of one contributor, and movement of a pairwise ranking gap. The exact aggregate-set construction is useful for proofs and numerical evaluation, but is not needed to understand this interpretation; it is given in \cref{app:geometry-corollaries,app:proof-shapley-geometry}.

As a numerical verification of this characterization, we evaluate the identity in \cref{thm:valuation-shift-body} on a finite three-contributor binary decision system. For 29 fixed directions $\xi$, the maximal directional Shapley change predicted from the accessible-output sets is compared with the value achieved by the downstream task recovered from the corresponding optimization. The maximum absolute discrepancy is $1.34\times10^{-10}$, and 5000 additional random tasks per direction never exceed the predicted bound. The full experimental details and verification plots are provided in \cref{app:geometry-verification}.

\subsection{A concrete two-contributor reversal}
\label{sec:reversal}

We now make the ranking reversal completely explicit. There are two contributors, $A$ and $B$, and one hidden class
\[
\Theta\in\{0,1,2\},
\quad
\Pr(\Theta=j)=\frac13.
\]
The same hidden class controls both the quantum training data and the downstream prediction task. If $\Theta=j$, contributor $A$ supplies a state $\rho_{j,A}$ and contributor $B$ supplies a state $\rho_{j,B}$. Thus the two contributors provide different physical encodings of the same unknown quantity $\Theta$.

The learner first measures whichever contributor states are available. A measurement produces a classical outcome, say $Z=z$, whose distribution depends on $\Theta$ because the measured quantum state depends on $\Theta$. From this outcome, the learner chooses an estimate $\widehat\Theta$. For the optimal strategy, it chooses the value of $\Theta$ that is most likely given the observed outcome $z$. For a contributor subset $S$ and access model $\cR$, we write
\[
p_{\cR}(S)
\]
for the largest probability of obtaining the correct estimate, $\Pr(\widehat\Theta=\Theta)$, when both the measurement and the final decision rule are optimized subject to $\cR$.

The estimate $\widehat\Theta$ is then used in a simple supervised test task. The test input is drawn uniformly from $X\in\{1,2,3\}$, and the hidden class selects the true input--output rule
\[
\begin{array}{c|ccc}
 & X=1 & X=2 & X=3\\
\hline
h_0 & 0 & 0 & 0\\
h_1 & 0 & 1 & 1\\
h_2 & 1 & 0 & 1
\end{array}
\]
That is, if $\Theta=j$, then the true label is
\[
Y=h_j(X),
\]
while the learner predicts
\[
\widehat Y=h_{\widehat\Theta}(X).
\]
For example, if $\Theta=1$, then the true rule is $h_1$: the correct labels at $X=1,2,3$ are respectively $0,1,1$.

The three rules are chosen so that any two of them disagree on exactly two of the three possible test inputs. Therefore, if the quantum measurement leads to the correct class estimate $\widehat\Theta=\Theta$, the learner chooses the correct rule and has zero test error. If it chooses the wrong class, it uses one of the other two rules and has test error $2/3$. Hence
\begin{equation}
R_{\cR}^{\star}(S)
=
\frac23\left[1-p_{\cR}(S)\right].
\label{eq:risk-success}
\end{equation}
With no contributor data, the best strategy is to guess one of the three classes, so $p(\varnothing)=1/3$ and $R_{\emptyset}=4/9$. Using $v_{\cR}^{\star}(S)=R_{\emptyset}-R_{\cR}^{\star}(S)$ then gives
\begin{equation}
v_{\cR}^{\star}(S)
=
\frac23\left[p_{\cR}(S)-\frac13\right].
\label{eq:value-success}
\end{equation}
Thus, in this example, a contributor is valuable precisely to the extent that its quantum data help the learner identify $\Theta$, and therefore select the correct prediction rule.

Contributor $A$ is deliberately simple and access-independent. It supplies a four-level classical register,
\begin{equation}
\rho_{\theta,A}
=
s\,\ketbra{\theta}{\theta}
+
(1-s)\ketbra{e}{e},
\quad
s=\frac{37}{40},
\label{eq:erasure-contributor-A}
\end{equation}
where $\ket e$ is an erasure symbol orthogonal to $\ket0,\ket1,\ket2$. Measuring this register either reveals $\theta$ exactly, with probability $s$, or returns the erasure symbol. After an erasure the learner has no remaining information from $A$ and guesses uniformly. Hence
\begin{equation}
p_A
=
s+\frac{1-s}{3}
=
\frac{19}{20}
=
0.95.
\label{eq:pA}
\end{equation}
Because $A$ is already classical, this success probability is the same under both access models.

Contributor $B$ carries genuinely quantum information. For $\Theta=j$, it supplies the two-qubit product state
\[
\ket{D_j}
=
\ket{s_j}\otimes\ket{s_j},
\]
where
\begin{align}
\ket{s_0}&=\ket0,\\
\ket{s_1}&=-\frac12\ket0+\frac{\sqrt3}{2}\ket1,\\
\ket{s_2}&=-\frac12\ket0-\frac{\sqrt3}{2}\ket1.
\end{align}
These are the three double-trine states. Each $\ket{D_j}$ is a product state, but the optimal discrimination probability depends on how the two qubits may be measured \citep{peres1991optimal,chitambar2013return}. Under one-way LOCC, denoted $\cR_{\LOCC}$, the optimal correct-identification probability is 
\begin{equation}
p_B^{\LOCC}
=
\frac12+\frac{\sqrt3}{4}
\approx0.9330,
\end{equation}
whereas an unrestricted joint measurement, denoted $\cR_G$, achieves
\begin{equation}
p_B^{G}
=
\frac12+\frac{\sqrt2}{3}
\approx0.9714.
\label{eq:doubletrine-optima}
\end{equation}
Therefore
\begin{equation}
p_B^{\LOCC}<p_A<p_B^{G}.
\label{eq:reversal-ordering}
\end{equation}
Using $B$ alone, the learner is less likely than with $A$ to identify $\Theta$ under one-way LOCC, but more likely than with $A$ under global access.

To compute actual Shapley values, we must also specify what happens when both contributors are available. We take their joint state to be conditionally independent given $\Theta$,
\[
\rho_{\theta,AB}
=
\rho_{\theta,A}\otimes\ketbra{D_\theta}{D_\theta}.
\]
The role of the two contributors is then transparent. The learner first reads the classical register from $A$. If $A$ reveals $\theta$, the class is known exactly. If $A$ returns the erasure symbol, the learner falls back on the quantum data from $B$. Because the erasure event is independent of $\Theta$,
\begin{equation}
p_{\cR}(AB)
=
s+(1-s)p_B^{\cR}.
\label{eq:joint-success-example}
\end{equation}

The resulting singleton and joint values are
\begin{align}
v(A)
&=
\frac{37}{90},\\
v_{\LOCC}(B)
&=
\frac19+\frac{\sqrt3}{6},\quad
v_G(B)=
\frac{1+2\sqrt2}{9},\\
v_{\LOCC}(AB)
&=
\frac{151}{360}+\frac{\sqrt3}{80},\quad
v_G(AB)=
\frac{151}{360}+\frac{\sqrt2}{60}.
\label{eq:example-coalition-values}
\end{align}

For two contributors and $v(\varnothing)=0$, the Shapley values are
\begin{align}
\phi_A
=&
\frac12v(A)+\frac12\left[v(AB)-v(B)\right],
\\
\phi_B
=&
\frac12v(B)+\frac12\left[v(AB)-v(A)\right].
\label{eq:two-player-shapley-example}
\end{align}
Substituting the values above gives, under one-way LOCC,
\begin{align}
\phi_A^{\LOCC}
=&
\frac{259}{720}-\frac{37\sqrt3}{480}
\approx0.2262,\\
\phi_B^{\LOCC}
=&
\frac{43}{720}+\frac{43\sqrt3}{480}
\approx0.2149,
\label{eq:example-shapley-locc}
\end{align}
so $A$ is ranked above $B$. Under global access,
\begin{align}
\phi_A^{G}
=&
\frac{259}{720}-\frac{37\sqrt2}{360}
\approx0.2144,
\\
\phi_B^{G}
=&
\frac{43}{720}+\frac{43\sqrt2}{360}
\approx0.2286,
\label{eq:example-shapley-global}
\end{align}
so the ranking reverses:
\begin{equation}
\phi_A^{\LOCC}>\phi_B^{\LOCC}
\quad\text{but}\quad
\phi_A^{G}<\phi_B^{G}.
\label{eq:one-copy-ranking-reversal}
\end{equation}

This is the full mechanism in numbers. Contributor $A$ does not change at all when the access model changes. Contributor $B$ also supplies exactly the same product states in both cases. The only change is whether the two qubits supplied by $B$ must be read by one-way LOCC or may be measured jointly. That change is enough to move $B$ from below $A$ to above $A$ in the final Shapley ranking. The derivation is collected in \cref{app:one-copy-construction}.

\section{From Best-Achievable Value to a Fixed Learner}
\label{sec:operational}

So far, $v_{\cR}^{\star}$ has meant the \emph{best-achievable value} under an access model: for each contributor subset, it asks for the best performance achievable by any learner allowed by the physical access model. That is the right object when the goal is to compare physical access models themselves. In practice, however, an experimenter may already have chosen one concrete training-and-prediction pipeline and may want to value contributors for that fixed learner. Once the learner is fixed, there is no longer an optimization over all protocols allowed by the access model. For each contributor subset, the same measurement, classical processing, and test procedure are applied every time. By the linearity of quantum measurement probabilities in the input density operator, the expected performance of this fixed pipeline can therefore be written as
\[
\Tr(F_S\varrho_S)
\]
for some operator $F_S$. This linear form is what later allows the
Shapley value itself to be represented as the expectation of a single
Shapley observable.

The notation changes slightly here for a reason. Earlier, $\rho_{\theta,S}$ described a family of possible states indexed by the unknown task parameter $\theta$. In this section we instead study one concrete finite-copy training block consumed by a fixed learner, so we denote its density operator by $\varrho_D$. We assume the block has identifiable contributor subsystems; when only contributors in $S$ are retained, the corresponding reduced training block is
\[
\varrho_S=\Tr_{S^c}(\varrho_D).
\]

\begin{definition}[Fixed-learner subset value]
\label{def:operational-value}
Fix one learning rule $\cA$ before performing the valuation, and let $R_{\cA}(S)$ be its expected test risk when trained using contributor subset $S$. Define
\begin{equation}
v_{\cA}(S)=R_{\cA}(\varnothing)-R_{\cA}(S),
\quad
\phi_i^{\cA}=\Sh_i(v_{\cA}).
\label{eq:operational-value}
\end{equation}
\end{definition}
This fixed-learner value answers a different question from the
best-achievable value $v_{\cR}^{\star}$: it measures the value actually realized by one specified learner rather than the best value available within the access model. 

Why does fixing the learner simplify the quantum problem? For a fixed subset $S$, the complete pipeline---quantum training protocol, classical model selection, and bounded test score---is now fixed. Its expected bounded test score is therefore a linear functional of the input density operator. In quantum information, such a bounded probability-valued linear functional is represented by an \emph{effect}, meaning a positive operator bounded by the identity. Hence there exists $0\le F_S\le\1_S$ whose expectation equals the bounded test score. To compare different subsets inside one common training block, we embed the effect into the full training space as 
\[
\widetilde F_S:=F_S\otimes\1_{S^c}.
\]
The Shapley-weighted difference of these effects defines one \emph{Shapley observable} for each contributor.

\begin{theorem}[Shapley observable representation]
\label{thm:valuation-observable}
For every fixed learner $\cA$ and contributor $i$,
\begin{align}
\phi_i^{\cA}&=\Tr(\Omega_i\varrho_D),\\
\Omega_i&=\sum_{S\subseteq N\setminus\{i\}}w_S
\big(\widetilde F_{S\cup\{i\}}-\widetilde F_S\big),\quad
\|\Omega_i\|_\infty\le1.
\label{eq:valuation-observable}
\end{align}
\end{theorem}

The interpretation is direct. Once the learner has been fixed, contributor $i$'s Shapley value is the expectation value of one Hermitian operator $\Omega_i$ on the complete finite-copy training block. This separates two jobs that are easy to conflate. The first is computational: construct or approximate the subset effects that define $\Omega_i$. The second is physical: estimate the expectation $\Tr(\Omega_i\varrho_D)$ from fresh preparations of the unknown training block. 

This representation also gives a direct physical route to estimating the Shapley value. Once the Shapley observable $\Omega_i$ has been constructed, $\phi_i^{\cA}$ can be estimated from repeated preparations of the training state using standard quantum expectation-estimation methods. Detailed sample complexity bounds and simultaneous estimation of multiple contributors are given in \cref{prop:single-copy-complexity,app:shadow-reduction}.

\section{EMPIRICAL EVIDENCE}
\label{sec:experiments}

The main theoretical results in \cref{thm:valuation-deficiency-duality,thm:valuation-shift-body} are exact, so the numerical experiments are not intended to prove them. Instead, the empirical studies address three practical questions. First, is the access-induced ranking reversal in \cref{sec:reversal} specific to a carefully constructed example, or does it also appear in nearby random quantum datasets? Second, does access dependence remain visible in a controlled finite-copy learning pipeline when the supplied quantum states, extracted features, downstream learner, and contributor subsets are fixed? Third, does the same phenomenon persist when the learner itself is quantum, so that the access restriction is embedded directly into the pooling operations of a quantum convolutional neural network (QCNN)? The numerical verification of the extremal Shapley-shift geometry, together with full protocols and additional diagnostics, is deferred to Appendix~\ref{app:experimental-details}.

\subsection{Access-induced ranking reversals in random product-state data}
\label{sec:exp-reversal}

The first experiment studies whether access-induced changes in Shapley attribution appear only in specially constructed examples or also occur in random quantum datasets. We generate a random quantum state-discrimination task with four contributors \citep{barnett2009discrimination,bae2015discrimination}. A hidden class
\[
T\in\{1,2,3\}
\]
is sampled with a uniform prior, and each contributor supplies a qubit state that depends on the same hidden class,
\[
\rho_{T,i},\quad i=1,\ldots,4 .
\]
For a coalition $S$, the learner receives the corresponding joint state
\[
\rho_{T,S}=\bigotimes_{i\in S}\rho_{T,i}
\]
and attempts to infer the hidden class $T$.

Let $p_{\cR}(S)$ denote the probability of correctly identifying $T$ under access model $\cR$. We use the normalized value
\[
v_{\cR}(S)=\frac23\left(p_{\cR}(S)-\frac13\right),
\]
so that random guessing has zero value. The Shapley value is then computed from the exact coalition values for all contributor subsets.

We compare two physical access models. Weak access performs local
product-Pauli measurements and applies a maximum-a-posteriori (MAP) decision rule to the classical outcomes. Strong access allows a global measurement on the complete coalition state. The only difference between the two settings is the allowed measurement access; the contributor states, coalition structure, and downstream decision problem remain unchanged.

For each class $T\in\{1,2,3\}$ and contributor $i\in\{1,\ldots,4\}$,
we independently draw a Haar-random pure qubit
\[
|\psi_{T,i}\rangle
=
\alpha_{T,i}|0\rangle+\beta_{T,i}|1\rangle,
\quad
|\alpha_{T,i}|^2+|\beta_{T,i}|^2=1,
\]
and set $\rho^{\rm H}_{T,i}=|\psi_{T,i}\rangle\!\langle\psi_{T,i}|$. We then define
\begin{equation}
\rho_{T,i}^{(\kappa,\nu)}
=(1-\nu)\left[(1-\kappa)\mathcal D_Z(\rho^{\rm H}_{T,i})
+\kappa\rho^{\rm H}_{T,i}\right]
+\nu\frac{I_2}{2}.
\label{eq:random-state-generator}
\end{equation}
where $\mathcal D_Z$ removes the off-diagonal entries in the computational basis. Thus $\kappa$ scales only the computational-basis coherence, while $\nu$ applies standard qubit depolarization. For each dataset seed, the same 12 Haar projectors are reused across $\kappa\in\{0,.25,.50,.75,1\}$ and $\nu\in\{0,.1,.2,.3\}$.

At $\kappa=0$, every $\rho_{T,i}^{(0,\nu)}$ is diagonal in the same
computational basis, so all coalition states commute. Because the Weak product-Pauli library contains the all-$Z$ measurement followed by MAP decoding, it attains the globally optimal discrimination performance in this limit; Strong access cannot improve on it. Increasing $\kappa$ restores transverse coherence, making the hypotheses generically noncommuting and creating room for the two access models to extract different information from the same contributor states.

\begin{table}[htbp]
\centering
\caption{\textbf{Random product-state reversal summary.} The upper block reports ranking reversal fraction averaged across noise levels. The lower block uses all 65 baseline reversal instances at $(\kappa,\nu)=(1,.1)$, selected before perturbation; brackets are 95\% instance-bootstrap intervals from 20,000 resamples.}
\label{tab:random-reversal-summary}
\small
\setlength{\tabcolsep}{5.0pt}
\begin{tabular}{lccccc}
\toprule
Coherence $\kappa$ & $0$ & $.25$ & $.50$ & $.75$ & $1.00$ \\
\midrule
Reversal prevalence & 0.000 & 0.180 & 0.320 & 0.603 & 0.663 \\
\bottomrule
\end{tabular}
\vspace{4pt}

\begin{tabular}{lcc}
\toprule
Perturbation $\sigma$ & Any reversal & Same original pair\\
\midrule
$.01$ & $0.988\;[0.973,0.998]$ & $0.973\;[0.956,0.987]$ \\
$.03$ & $0.943\;[0.913,0.968]$ & $0.898\;[0.865,0.928]$ \\
$.05$ & $0.898\;[0.864,0.931]$ & $0.816\;[0.779,0.852]$ \\
$.10$ & $0.814\;[0.773,0.851]$ & $0.636\;[0.597,0.674]$ \\
\bottomrule
\end{tabular}
\end{table}

\begin{figure}[htbp] 
\centering \includegraphics[width=0.94\linewidth]{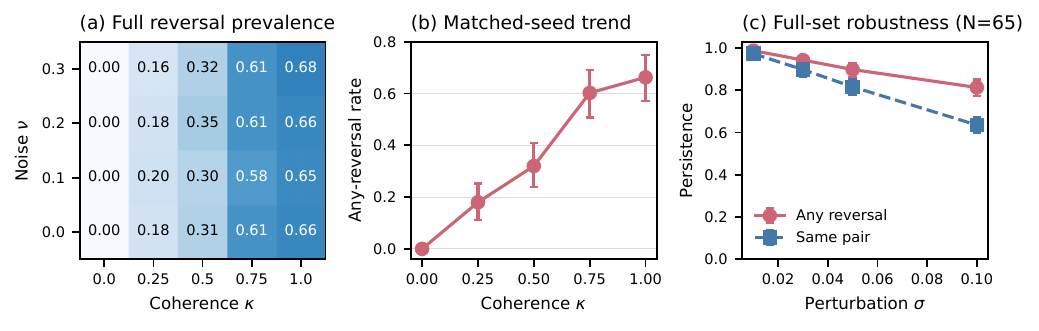} \caption{\textbf{Access-induced ranking reversals and their robustness.}
(a) Reversal prevalence across coherence $\kappa$ and depolarizing noise $\nu$.
(b) The matched-seed trend as coherence increases.
(c) Persistence of the 65 baseline reversal instances under increasing perturbation strength $\sigma$, reporting both retention of any reversal and retention of the same reversed contributor pair. Error bars represent 95\% instance-bootstrap confidence intervals.}
\label{fig:generic-reversal}
\end{figure}

The numerical results are summarized in
\cref{fig:generic-reversal,tab:random-reversal-summary}. The ranking reversal fraction in \cref{tab:random-reversal-summary} shows that access-induced Shapley reversals become increasingly frequent as the quantum coherence parameter $\kappa$ increases. In the strictly commuting limit $\kappa=0$, no reversal is observed, consistent with the equality of the optimal Weak and Strong discrimination performance. At $\kappa=1$, around two thirds of the generated datasets exhibit at least one contributor ranking reversal across all noise levels.

To examine whether these ranking reversals are isolated numerical effects, we perturb the baseline reversal instances and measure their persistence. As shown in \cref{fig:generic-reversal}, most reversals remain after perturbation, and a substantial fraction preserve the same contributor pair. This indicates that access-induced ranking changes are not caused by accidental near ties, but persist over finite neighborhoods of the generated quantum-data space.

\subsection{Physical access changes Shapley attribution in a controlled TFIM learning task}
\label{sec:exp-tfim}

The previous experiments show that changing the allowed quantum access can change contributor values in abstract decision problems. We now test whether the same phenomenon appears in a controlled quantum-learning pipeline. The goal is to determine whether the same quantum training examples receive different Shapley values and rankings when only the physical procedure used to extract information from them is changed. 

We consider a binary classification task where the goal is to predict a physical phase label from quantum states. Each example is generated from a six-site transverse-field Ising model (TFIM). 
TFIM is controlled by a physical parameter $g$: for each value of $g$, the model generates a quantum state 
\[
|\psi(g)\rangle ,
\]
which is the ground state of the corresponding Hamiltonian
\[
H(g)=
-\sum_{j=1}^{5}Z_jZ_{j+1}
-g\sum_{j=1}^{6}X_j .
\]

The learner cannot directly use quantum states as classical inputs. We therefore represent each quantum state using a fixed set of eight fidelity features \citep{schuld2019feature}. Specifically, we choose eight reference TFIM ground states with parameters
\[
g_r^{\rm ref}\in
\{0.35,0.55,0.75,0.95,1.05,1.25,1.45,1.65\}.
\]
For a contributor state $|\psi(g_i)\rangle$, the $r$-th feature is defined as the fidelity with the $r$-th reference state, 
\[
F_r(g_i)=
|\langle\psi(g_r^{\rm ref})|\psi(g_i)\rangle|^2 .
\]
The resulting feature vector is
\[
x_i=(F_1(g_i),\ldots,F_8(g_i)).
\]
All access models estimate these same eight fidelity coordinates. The only difference is the physical measurement procedure used to obtain them.

There are 10 contributors in total. A coalition of contributors provides the corresponding subset of training examples.

For each coalition, the resulting feature vectors are used to train the same classifier,
\[
\texttt{StandardScaler}+\texttt{LogisticRegression},
\]
and performance is evaluated on a fixed test set containing TFIM states from both sides of the transition. The coalition value is defined as the classification accuracy improvement over random guessing:
\[
v(S)=\operatorname{Accuracy}(S)-0.5 .
\]

Since there are 10 contributors, all possible coalitions ($2^{10}$ in total) are enumerated exactly. This allows the exact computation of the 10-dimensional Shapley vector for each physical access model.

We compare three finite-copy access models: uniform local Pauli shadows \citep{huang2020shadows}, observable-aware local Pauli measurements (OALP), and coherent reference-assisted SWAP measurements \citep{buhrman2001fingerprinting}. We also compute an exact-feature oracle using exact fidelity values as a reference. The oracle is a reference attribution for this fixed feature representation: it is obtained by replacing finite-copy feature estimates with exact fidelity coordinates while keeping the contributors, learner, and test procedure unchanged. It is not an intrinsic contributor value or a universal ground truth beyond this benchmark.

The main objects are therefore
\[
\phi_{\rm local},
\quad
\phi_{\rm OALP},
\quad
\phi_{\rm SWAP},
\quad
\phi_{\rm oracle}
\in\mathbb R^{10}.
\]
The primary question is whether different physical access models assign different values and rankings to the same contributors. The oracle-based comparisons are reported separately as secondary calibration metrics. 

\begin{figure}[htbp]
  \centering
  \includegraphics[width=0.95\linewidth]{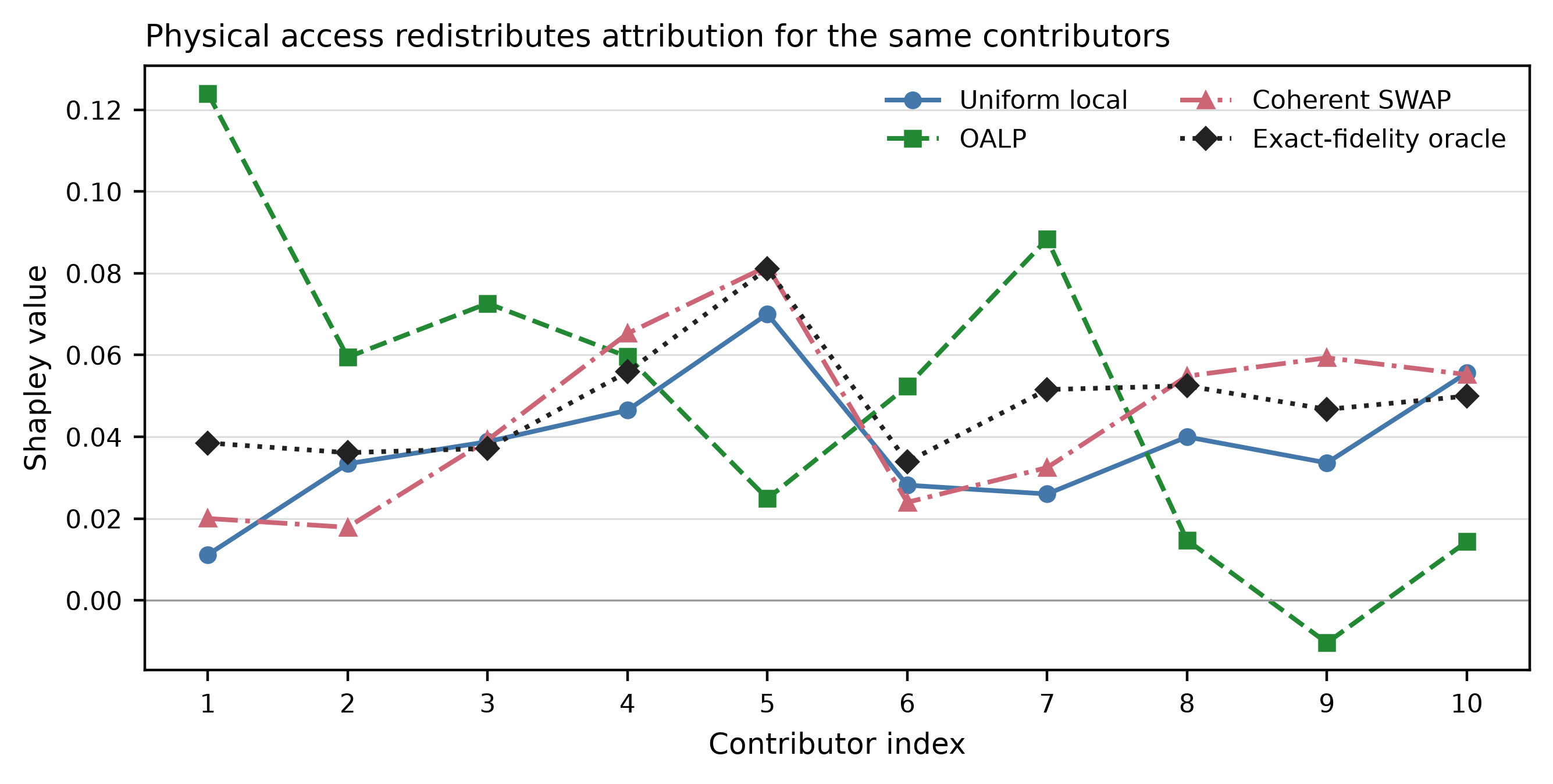}
  \caption{\textbf{Physical access redistributes attribution for the same contributors.} Shapley values of the 10 fixed TFIM contributors under uniform local Pauli access, observable-aware local Pauli access (OALP), coherent reference-assisted SWAP access, and the exact-feature oracle. All methods use the same contributors, quantum states, feature coordinates, learner, and coalition definitions; only the physical procedure used to obtain the features is changed.}
  \label{fig:tfim-shapley-attribution}
\end{figure}

As shown in \cref{fig:tfim-shapley-attribution}, changing only the physical access model changes the Shapley values and rankings assigned to the same quantum training samples, while the underlying states, feature representation, and downstream learner remain fixed. This directly demonstrates that quantum-data attribution is access-dependent. 

\subsection{Physical pooling access changes Shapley rankings in a QCNN}
\label{sec:exp-qcnn}

The preceding TFIM experiment isolates access dependence in a pipeline where quantum states are first converted into fidelity features and then processed by a classical classifier. We next ask whether the same contributor-level effect persists when the learner itself is quantum and the access restriction is part of the network \citep{cong2019qcnn}. We use the same six-site TFIM phase-classification setting, with 10 quantum training contributors per dataset and the same balanced fixed test task. Contributor $i$ supplies a TFIM state $\rho_i=\ketbra{\psi(g_i)}{\psi(g_i)}$ and its phase label. For each coalition $S\subseteq N$ and each access rule $R\in\{\mathrm W,\mathrm S\}$, the QCNN is trained independently on the contributors in $S$ and evaluated on the fixed test set. Writing $A_R(S)$ for the expected probability of a correct binary output, we define
\begin{equation}
  v_R(S)=A_R(S)-\frac12,
  \qquad
  \phi_i^R=\Sh_i(v_R).
  \label{eq:qcnn-coalition-value}
\end{equation}
Thus the fixed learner is the prescribed architecture and training rule, not a single fitted parameter vector: every trainable coalition is retrained from its matched initialization. All $2^{10}$ coalitions are enumerated exactly.

The QCNN compresses 6 input qubits to 3 and then to 1 output qubit. The Weak and Strong models share the same convolutional architecture and differ only in the pooling access allowed on a source--target pair $(A,B)$. Weak pooling measures the source after a trainable single-qubit basis rotation and applies an outcome-conditioned single-qubit unitary to the retained target, 
\begin{equation}
  \mathcal P_{\mathrm W}(\rho_{AB})
  =\sum_m V_m\,
  \Tr_A\!\left[(\Pi_m\otimes\1)\rho_{AB}(\Pi_m\otimes\1)\right]
  V_m^\dagger .
  \label{eq:qcnn-weak-pooling}
\end{equation}
Strong pooling contains all Weak operations and additionally permits coherent two-qubit preprocessing before the source measurement,
\begin{align}
  \mathcal P_{\mathrm S}(\rho_{AB})
  &=\sum_m V_m\,
  \Tr_A\!\left[(\Pi_m\otimes\1)W\rho_{AB}W^\dagger
  (\Pi_m\otimes\1)\right]V_m^\dagger,\\
  W&=R_{XX}(\gamma_x)R_{YY}(\gamma_y)R_{ZZ}(\gamma_z).
  \label{eq:qcnn-strong-pooling}
\end{align}
Setting $\boldsymbol\gamma=0$ recovers the Weak model exactly, so the Strong access class contains the Weak one. The overall $6\!\to\!3\!\to\!1$ network and the two pooling instruments are summarized schematically in \cref{fig:qcnn-schematic}. Detailed access-strictness and optimization audits are reported in Appendix~\ref{app:qcnn-details}.

\begin{figure}[htbp]
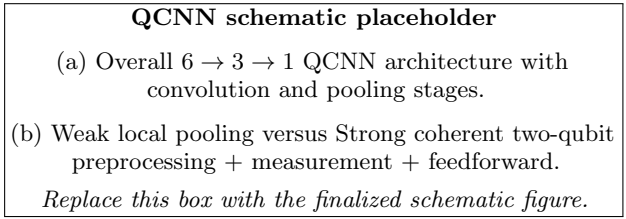

  \centering
  \fbox{\parbox[c][0.30\linewidth][c]{0.92\linewidth}{\centering
  \textbf{QCNN schematic placeholder}\\[6pt]
  (a) Overall $6\to3\to1$ QCNN architecture with convolution and pooling stages.\\[6pt]
  (b) Weak local pooling versus Strong coherent two-qubit preprocessing + measurement + feedforward.\\[4pt]
  \emph{Replace this box with the finalized schematic figure.}}}
  \caption{\textbf{QCNN architecture and physical pooling access.}
  The shared $6\to3\to1$ network differs only in the pooling access. Weak pooling measures the source locally before classical feedforward to the retained qubit, whereas Strong pooling additionally permits coherent two-qubit preprocessing before that measurement.}
  \label{fig:qcnn-schematic}
\end{figure}

\begin{figure}[htbp]
  \centering
  \includegraphics[width=\linewidth]{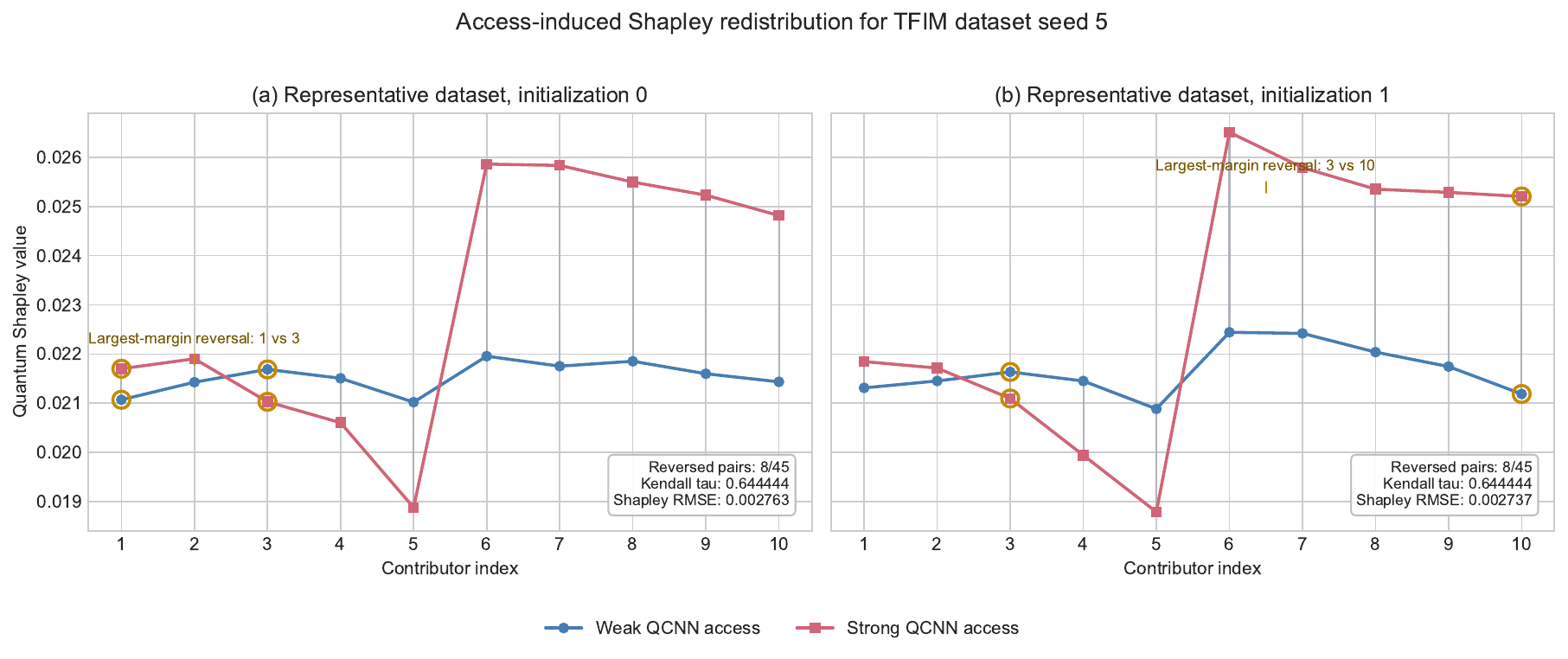}
  \caption{\textbf{Access-induced Shapley redistribution in a representative QCNN dataset.} Weak and Strong QCNN Shapley values for TFIM dataset seed 5 under two matched initialization schedules. The highlighted pair in each panel is the largest-margin ranking reversal for that initialization.}
  \label{fig:qcnn-seed5-shapley}
\end{figure}

\begin{figure}[htbp]
  \centering
  \includegraphics[width=\linewidth]{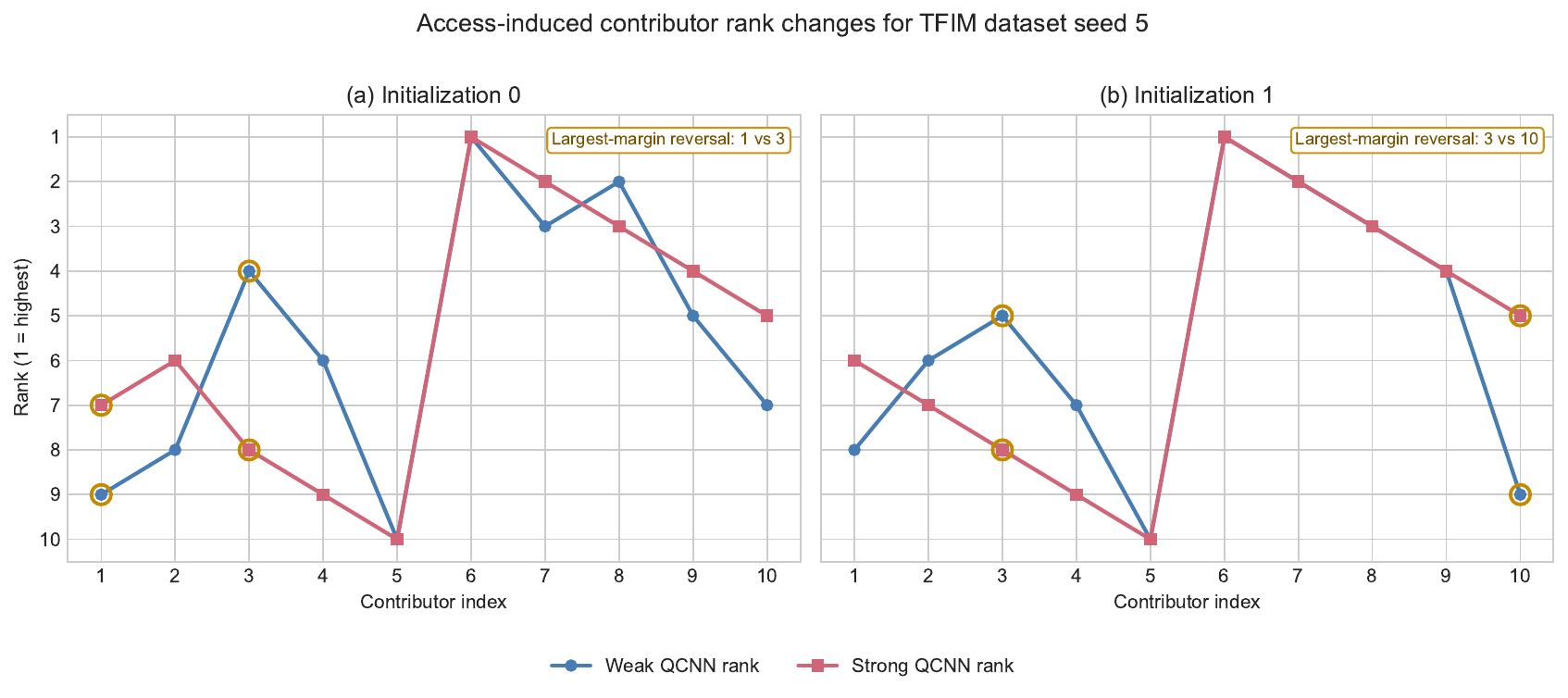}
  \caption{\textbf{Physical pooling access changes contributor rankings.}
  The same seed-5 Shapley vectors are converted to contributor ranks, with rank one placed at the top. Crossings between the Weak and Strong rank curves make the access-induced ordering changes explicit under both matched initializations.}
  \label{fig:qcnn-seed5-rank}
\end{figure}

We first visualize a representative dataset (seed 5). This seed is not an extremal example: its pairwise reversal fraction and Kendall rank agreement coincide with the across-dataset medians under both matched initialization schedules. As shown in \cref{fig:qcnn-seed5-shapley}, changing only the pooling access redistributes the 10 Shapley values. The induced shifts are large enough to change contributor orderings, as displayed directly in \cref{fig:qcnn-seed5-rank}. Under both initializations, changing the pooling access produces a clear redistribution of contributor ranks, including multiple pairwise order reversals in this representative dataset.

\begin{figure}[htbp]
  \centering
  \includegraphics[width=\linewidth]{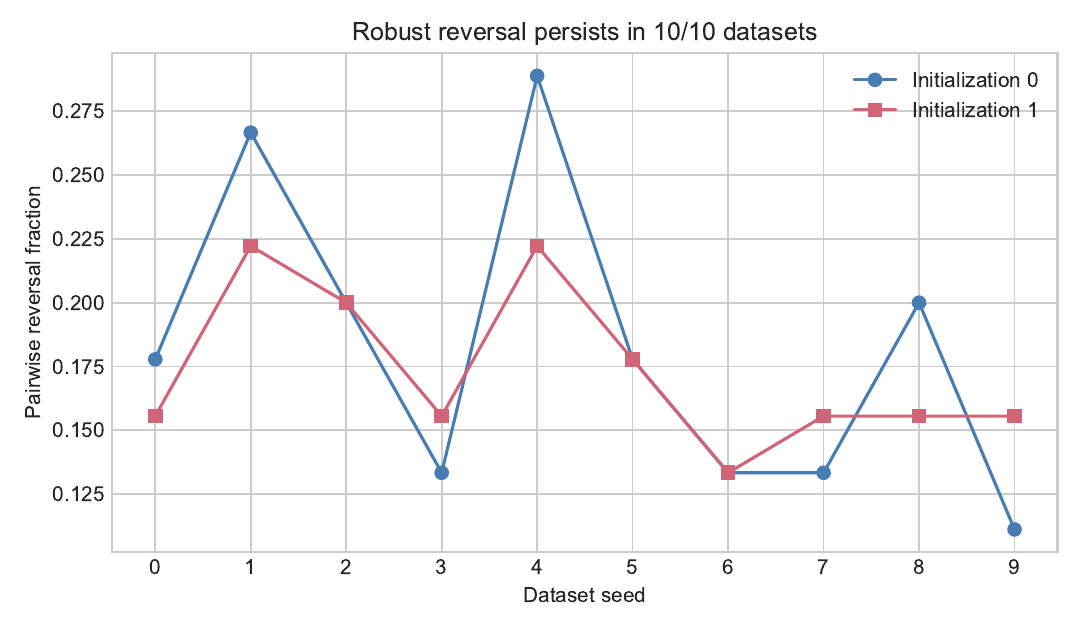}
  \caption{\textbf{Access-induced ranking reversals persist across datasets and initializations.} Pairwise Weak--Strong reversal fractions for all 10 pre-specified TFIM datasets under two matched QCNN initialization schedules. 
  Every dataset contains at least one non-near-tie reversal under each initialization.}
  \label{fig:qcnn-reversal-fraction}
\end{figure}

The representative behavior persists across independently generated training sets. With the QCNN architecture, optimizer, exact simulator, coalition rules, and access definitions frozen, all 10 pre-specified TFIM datasets exhibit at least one non-near-tie Weak--Strong ranking reversal under the first initialization schedule, and all 10 do so again under the second. The mean pairwise reversal fractions are $0.182$ and $0.173$, respectively, while the mean Kendall agreements are $0.636$ and $0.653$. The seed-wise reversal fractions are shown in \cref{fig:qcnn-reversal-fraction}. 

To separate the access effect from ordinary optimization variability, we also compare the displacement of the full 10-dimensional Shapley vector. For dataset $d$ and initialization $r$, define 
\begin{equation}
 D_{\rm access}^{(d,r)}
 =\left[\frac1{10}\sum_{i=1}^{10}
 \big(\phi_{d,r,i}^{\mathrm S}-\phi_{d,r,i}^{\mathrm W}\big)^2\right]^{1/2},
 \label{eq:qcnn-access-rmse}
\end{equation}
and, at fixed access $R$, define the cross-initialization displacement
\begin{equation}
 D_{{\rm init},R}^{(d)}
 =\left[\frac1{10}\sum_{i=1}^{10}
 \big(\phi_{d,1,i}^{R}-\phi_{d,0,i}^{R}\big)^2\right]^{1/2}.
 \label{eq:qcnn-init-rmse}
\end{equation}
Across the 10 datasets, the mean access-induced RMSE is
$2.83\times10^{-3}$ for initialization 0 and $2.75\times10^{-3}$ for
initialization 1, whereas the mean cross-initialization RMSE is only
$3.03\times10^{-4}$ within Weak access and $3.21\times10^{-4}$ within Strong access. 
Thus the valuation shift produced by changing physical pooling access is about an order of magnitude larger than the shift produced by changing initialization alone. 

As an implementation check, the trained Strong QCNNs consistently make nontrivial use of the enlarged pooling access: every trained Strong coalition exhibits a non-product effective pooling effect. Further access-strictness, pair-persistence, coalition-level, training, and Shapley-consistency diagnostics are reported in Appendix~\ref{app:qcnn-details}.

\section{Discussion and Conclusion}
\label{sec:discussion}

The central conclusion of this work is that quantum data value is operational rather than intrinsic to the supplied states alone. For finite-copy quantum data, the information available to a learner depends on how those states may be physically accessed. Consequently, even when the supplied quantum states and downstream task are fixed, changing the admissible physical access can change coalition values and, in turn, the values and rankings assigned to individual data contributors. An access-aware valuation therefore describes the interaction among the quantum states, the physical means by which information is extracted from them, and the downstream task. 

Our theoretical results make this dependence quantitative. For nested access models, \cref{thm:valuation-deficiency-duality} links the operational advantage of richer access exactly to Shapley ranking reversal, while \cref{thm:valuation-shift-body} extends this picture to the full attribution vector for arbitrary pairs of access models. For a fixed learning rule, \cref{thm:valuation-observable} further shows that each realized Shapley value can be represented as the expectation of a Hermitian observable on the finite-copy training block. Together, these results connect physical access, task utility, contributor-level attribution, and finite-copy estimation within a common framework. 

The experiments show that this access dependence persists beyond the theoretical constructions. It appears in random quantum discrimination tasks, in a controlled TFIM pipeline where only the physical feature-extraction procedure is changed, and in a quantum-native QCNN where the access restriction is built directly into the pooling operation. In each case, changing physical access alone can alter the Shapley values and rankings of data contributors. This implies that quantum-data selection and measurement design cannot, in general, be treated as independent stages: which data are most valuable may depend on the physical operations available when those data are used. 

Several directions follow naturally. One is the joint design of data acquisition and physical access, where data and measurement resources are optimized together rather than sequentially. Another is cost-aware valuation, which would balance access-induced gains in data value against the physical cost of collective measurements, coherent control, ancillary systems, or additional state preparations. It is also natural to study how access-dependent attribution evolves with copy budget and across broader hierarchies of local, adaptive, collective, and architecture-dependent access.

\section*{Data Availability Statement}
The code and numerical data supporting the findings of this study will be made publicly available upon publication. Reproducibility settings and additional numerical diagnostics are provided in the appendices.

\clearpage
\appendix

\section{Experimental Details and Additional Results}
\label{app:experimental-details}

\subsection{Secondary TFIM attribution calibration}
\label{app:tfim-attribution-calibration}

The following tables provide oracle-relative calibration metrics and direct
pairwise attribution comparisons. They are secondary analyses; the primary
observation is the Shapley redistribution shown in
\cref{fig:tfim-shapley-attribution}.

\begin{table}[htbp]
\centering
\caption{\textbf{Contributor attribution under different physical access models at $B=256$.}
All methods use the same contributors, quantum states, feature coordinates,
classifier, coalitions, and test set.}
\label{tab:tfim-attribution-oracle}
\small
\begin{tabular}{lccc}
\toprule
Access model & Kendall $\uparrow$ &
Shapley RMSE $\downarrow$ & Overlap@3 $\uparrow$ \\
\midrule
Uniform local Pauli & 0.102 & 0.0268 & 0.327 \\
OALP & 0.146 & 0.0240 & 0.357 \\
Coherent SWAP & 0.602 & 0.0117 & 0.650 \\
\bottomrule
\end{tabular}
\end{table}

\begin{table}[htbp]
\centering
\caption{\textbf{Direct access-induced Shapley changes at $B=256$.}
Kendall agreement compares contributor rankings, while pairwise reversal
fraction measures how often two contributors exchange order.}
\label{tab:tfim-access-shift}
\small
\begin{tabular}{lcc}
\toprule
Comparison & Kendall agreement & Pairwise reversal \\
\midrule
Uniform local Pauli vs coherent SWAP & 0.125 & 0.437 \\
OALP vs coherent SWAP & 0.123 & 0.438 \\
\bottomrule
\end{tabular}
\end{table}

The first table compares each finite-copy attribution vector with the
exact-feature oracle. The second table directly compares attribution changes
between access models.

\subsection{Reproducibility and uncertainty}
All experiments use master seed $20260812$ with child streams generated by NumPy \texttt{SeedSequence}. Global discrimination uses SCS with tolerance $10^{-7}$ and at most $200{,}000$ iterations; primal--dual spot checks satisfy the pre-specified gates. Matched seeds, nested measurement streams, and matched access-model pairing are preserved throughout. For the pre-specified TFIM endpoint $N_{\mathrm{prep}}=256$, uncertainty for the three SWAP--OALP contrasts is computed at the dataset level: within each of 10 dataset seeds we average the 10 paired measurement-seed differences, then form a two-sided Student-$t$ 95\% interval across the resulting 10 block means. The corresponding intervals are $[0.384,0.528]$ for Kendall, $[-0.0146,-0.0099]$ for Shapley RMSE, and $[0.258,0.329]$ for Overlap@3; all 10 block means have the manuscript direction for each contrast. As a diagnostic, a 20,000-replicate dataset-cluster bootstrap gives $[0.393,0.509]$, $[-0.0142,-0.0104]$, and $[0.267,0.323]$, respectively. Full cell statistics and per-budget TFIM summaries are provided with the artifact.

\noindent\subsection{Expanded reversal-neighborhood audit}
At $(\kappa,\nu)=(1,.1)$, 65 of the fixed 100 baseline datasets exhibit at least one access-induced ranking reversal. These 65 instances were identified from the unperturbed baseline before any robustness outcomes were inspected. For each instance and each $\sigma\in\{.01,.03,.05,.10\}$, we generate 50 independent SU(2) perturbations using the same frozen state generator, Weak product-Pauli+MAP and Strong global-POVM pipelines, exact Shapley computation, tie tolerance, and SCS settings as in the original benchmark, yielding 13,000 perturbation trials in total. The statistical unit is the baseline dataset instance, and uncertainty in \cref{tab:random-reversal-summary} is obtained by 20,000 instance-level bootstrap resamples. No baseline instance was removed or rerun based on its perturbation outcome. The full-65 estimates closely track the original first-ten diagnostic: at $\sigma=.10$, any-reversal persistence increases from $0.788$ to $0.814$, while same-original-pair persistence changes from $0.638$ to $0.636$.

\begin{figure}[htbp]
  \centering
  \includegraphics[width=0.72\linewidth]{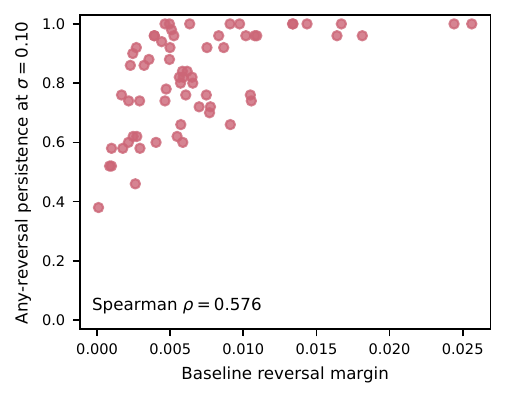}
  \caption{\textbf{Reversal margin predicts perturbation stability.} Baseline reversal margin versus persistence at $\sigma=.10$ across the 65 baseline reversal instances. Larger unperturbed reversal margins are associated with higher probability of retaining any reversal and, more strongly, the same original reversed pair.}
  \label{fig:margin-persistence}
\end{figure}

Robustness is also structured by the unperturbed reversal margin. Defining an instance margin as the maximum, over its originally reversed pairs, of the smaller absolute Weak/Strong pairwise gap, the margin correlates with $\sigma=.10$ persistence: Spearman $\rho=0.576$ for retaining any reversal ($p=5.30\times10^{-7}$) and $\rho=0.747$ for retaining the same original pair ($p=9.39\times10^{-13}$). Thus the least stable cases are disproportionately near-tie reversals rather than evidence that the access-induced effect disappears under perturbation; \cref{fig:margin-persistence} visualizes this diagnostic.

\begin{figure}[htbp]
  \centering
  \includegraphics[width=\linewidth]{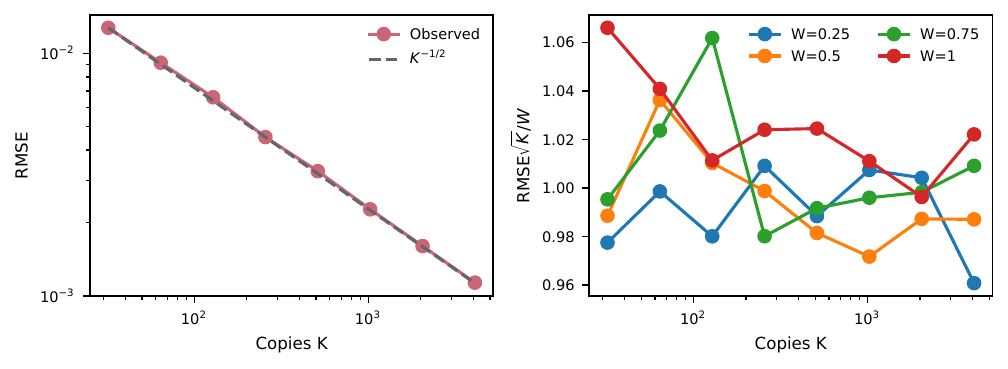}
  \caption{\textbf{Fixed-learner Shapley observable scaling.} Left: mean RMSE versus copies on log--log axes and a $K^{-1/2}$ reference; the fitted slope is $-0.501$. Right: the controlled-width diagnostic $\mathrm{RMSE}\sqrt K/W$ for $W\in\{.25,.5,.75,1\}$.}
  \label{fig:observable-scaling}
\end{figure}

\begin{figure}[htbp]
  \centering
  \includegraphics[width=\linewidth]{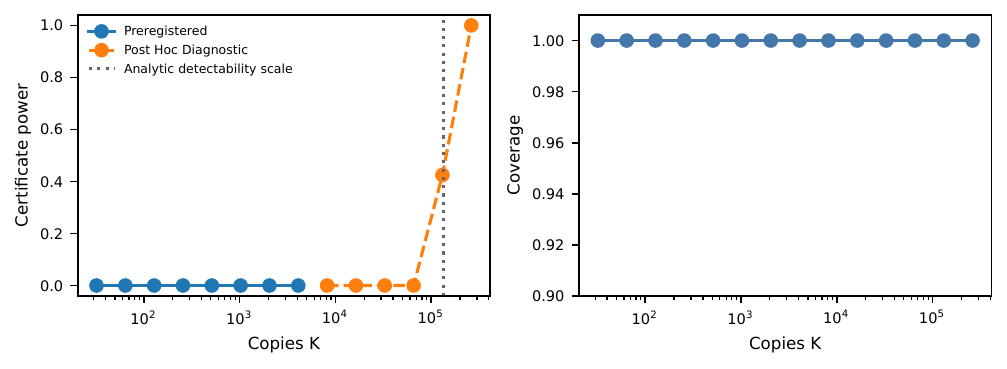}
  \caption{\textbf{Finite-copy resource certificate.} Pre-specified power is zero through $K=4096$ while coverage is one.  The visually separated post-hoc diagnostic begins to gain power at $K=131{,}072$; the vertical line is the analytic detectability scale $1.35\times10^5$, not a fitted threshold.}
  \label{fig:certificate-diagnostic}
\end{figure}

\begin{figure}[htbp]
  \centering
  \includegraphics[width=0.62\linewidth]{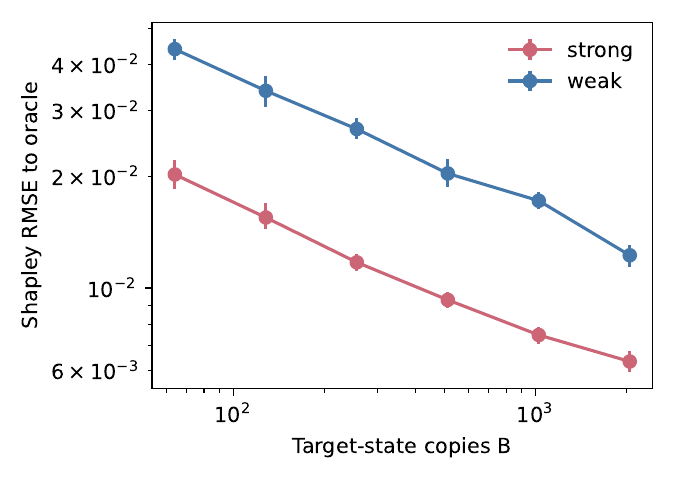}
  \caption{\textbf{Oracle-distance of TFIM Shapley values.} Shapley RMSE to the exact-feature oracle for the uniform local Pauli and coherent reference-assisted SWAP access pipelines versus target-state budget $N_{\mathrm{prep}}$.}
  \label{fig:tfim-oracle-error}
\end{figure}

\noindent\subsection{Controlled TFIM protocol and strengthened local control}
The test set contains 30 evenly spaced fields in $[.35,.95]$ and 30 in $[1.05,1.65]$, with references at $g=.35,.55,.75,.95,1.05,1.25,1.45,1.65$. The downstream learner is \texttt{StandardScaler}+\texttt{LogisticRegression} (\texttt{C=1}, \texttt{liblinear}, 5000 iterations); empty and one-class coalitions use the same constant rules as before. Uniform local Pauli access independently samples local $X/Y/Z$ bases and forms tensor-product shadow snapshots $(3\ketbra{b}{b}-\1)^{\otimes 6}$, averaging $N_{\mathrm{prep}}$ snapshots without clipping to estimate all eight fidelities. Coherent reference-assisted SWAP access estimates the same coordinates from $\pm1$ SWAP outcomes with deterministic round-robin allocation of the same $N_{\mathrm{prep}}$ target-state preparations. The exact-feature oracle uses exact fidelities. We evaluate $N_{\mathrm{prep}}\in\{64,128,256,512,1024,2048\}$ with 10 dataset seeds and 10 nested measurement seeds each, giving 100 paired replicates per budget. Here $N_{\mathrm{prep}}$ counts target-state preparations only. Reference-state preparations and coherent-control overhead required by SWAP access are held outside this copy-budget accounting, so the study compares access routes at matched target-copy budget rather than end-to-end physical cost.

For OALP, each reference projector $O_r=\ketbra{\psi_r}{\psi_r}$ is expanded in the six-qubit Pauli basis. Across the $3^6=729$ product-Pauli bases, we optimize a state-independent sampling distribution $q$ to minimize the maximum reference-wise inverse-coverage variance proxy $\max_r\sum_{P\ne I}c_{r,P}^2/p_P(q)$, subject to strictly positive coverage. Only the eight fixed reference projectors enter this optimization; no target state, label, exact target fidelity, oracle valuation, or downstream result is used. The optimized objective is $2.294$ versus $5.506$ under uniform sampling, the effective number of settings is $561.5$, and the maximum projector-reconstruction error is $3.83\times10^{-16}$. SCS returned \texttt{optimal}; all seeds completed without numerical instability. Existing E5 headline values were reproduced exactly from saved outputs before the new control was run.

Top-$k$ selection uses the Shapley vector from each paired replicate and deterministic contributor-index tie breaking. Overlap@$k$ is $|T_k^{\mathrm{method}}\cap T_k^{\mathrm{oracle}}|/k$. The pre-specified primary endpoint is $N_{\mathrm{prep}}=256$, $k=3$; $k=5,7$ are sensitivity analyses.

\begin{table}[htbp]
\centering
\caption{\textbf{TFIM contributor-selection sensitivity at $N_{\mathrm{prep}}=256$.} Exact-set match is intentionally stringent: any single membership error makes the replicate a mismatch.}
\label{tab:tfim-selection-sensitivity}
\small
\begin{tabular}{lcccc}
\toprule
$k$ & Access & Overlap & Jaccard & Exact-set match \\
\midrule
3 & Uniform local Pauli & 0.327 & 0.222 & 0.00 \\
3 & OALP & 0.357 & 0.247 & 0.02 \\
3 & Coherent SWAP & \textbf{0.650} & \textbf{0.505} & \textbf{0.10} \\
\midrule
5 & Uniform local Pauli & 0.566 & 0.421 & 0.02 \\
5 & OALP & 0.576 & 0.426 & 0.01 \\
5 & Coherent SWAP & \textbf{0.798} & \textbf{0.680} & \textbf{0.15} \\
\midrule
7 & Uniform local Pauli & 0.721 & 0.577 & 0.02 \\
7 & OALP & 0.746 & 0.607 & 0.04 \\
7 & Coherent SWAP & \textbf{0.847} & \textbf{0.744} & \textbf{0.13} \\
\bottomrule
\end{tabular}
\end{table}

\subsection{QCNN access and robustness diagnostics}
\label{app:qcnn-details}

The QCNN experiment in \cref{sec:exp-qcnn} uses the same frozen $6\to3\to1$ architecture for Weak and Strong access. Convolutional source--target edges are separated from the pooling pairs so that the Weak network cannot absorb the Strong premeasurement entangler into an unrestricted convolution on the same pair. In the Strong model, $W=R_{XX}(\gamma_x)R_{YY}(\gamma_y)R_{ZZ}(\gamma_z)$ is initialized at $\boldsymbol\gamma=0$; with all shared parameters matched, this reproduces the
Weak output exactly. For nonzero $\boldsymbol\gamma$, the corresponding pooling effect can have operator-Schmidt rank larger than one, while the Weak effect remains a product effect. Across both initialization schedules, every trained Strong coalition has a non-product effective pooling effect according to this audit.

All 10 pre-existing TFIM dataset seeds are used without selection. For each
dataset and each access, every one of the $2^{10}=1024$ coalition masks is
accounted for. Trainable coalitions use a fresh model and exactly 200 Adam steps
at learning rate $0.03$ with exact complex-valued density-matrix expectations;
empty and one-class coalitions use the frozen constant rule. There are no warm
starts or outcome-dependent retries. The two matched initialization schedules
use different mask-derived parameter initializations while remaining identical
across datasets for a fixed mask and schedule; Weak and Strong share the same
non-$\gamma$ initialization within each matched cell.

\begin{table}[htbp]
\centering
\caption{\textbf{QCNN exact-replication and initialization-robustness summary.}
The dataset seed is the independent statistical unit. Pairwise-reversal Jaccard
compares the exact reversed-pair sets between the two initialization schedules
within the same dataset.}
\label{tab:qcnn-robustness-summary}
\small
\begin{tabular}{lc}
\toprule
Quantity & Result \\
\midrule
Datasets with reversal, initialization 0 & $10/10$ \\
Datasets with reversal, initialization 1 & $10/10$ \\
Mean pairwise reversal fraction, init. 0 / 1 & $0.1822 / 0.1733$ \\
Mean Kendall agreement, init. 0 / 1 & $0.6356 / 0.6533$ \\
Mean access RMSE, init. 0 / 1 & $0.002833 / 0.002746$ \\
Mean cross-init RMSE, Weak / Strong & $0.000303 / 0.000321$ \\
Mean / median access-to-init RMSE ratio & $9.02 / 8.84$ \\
Mean reversed-pair Jaccard across initializations & $0.503$ \\
Strong coalitions with non-product pooling effect & $1.000$ \\
\bottomrule
\end{tabular}
\end{table}

The stronger pair-level persistence criterion is intentionally secondary. The
mean Jaccard overlap between the two reversed-pair sets is about $0.503$: the
existence of access-induced ranking reversal is stable across the two matched
initializations in all 10 datasets, while the identity of the particular
reversed pair can vary with optimization initialization. Exact Shapley
efficiency and the Shapley value of the access-difference game agree to below
$10^{-10}$ for every completed dataset. No dataset seed fails the training or
consistency checks.

\FloatBarrier

\section{Numerical verification of the Shapley shift geometry}
\label{app:geometry-verification}

This experiment verifies the directional identity in
\cref{thm:valuation-shift-body}. We use a fixed three-contributor binary
quantum decision system. The hidden state and action spaces are
\[
\theta\in\{0,1\},\quad a\in\{0,1\}.
\]
For each contributor $i$, the state family is generated independently, and a
coalition $S$ provides
\[
\rho_{\theta,S}=\bigotimes_{i\in S}\rho_{\theta,i}.
\]

For each coalition and access model, the allowed measurements and classical
decision rules define an accessible-output set
\[
K_{\cR}(S)\subset\mathbb R^2.
\]
A normalized binary decision task is represented by
\[
g=(g_0,g_1),\quad \|g\|_1\le 1,
\]
and the coalition value is the corresponding support function
\[
v_{\cR}(S;g)=h_{K_{\cR}(S)}(g).
\]

\begin{figure}[htbp]
  \centering
  \includegraphics[width=0.92\linewidth]{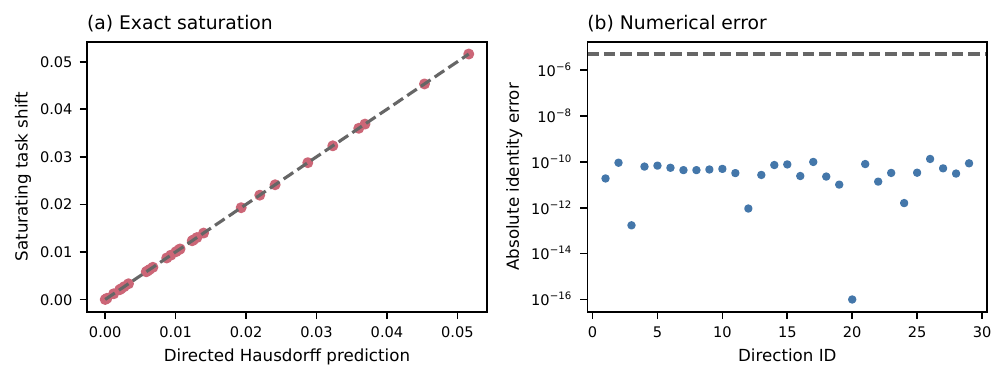}
  \caption{\textbf{Verification of maximal Shapley changes.} The predicted
  maximum directional Shapley shifts from the accessible-output construction
  agree with the shifts achieved by optimized downstream tasks for 29 fixed
  directions.}
  \label{fig:shift-geometry-exp}
\end{figure}

For each of 29 fixed directions $\xi$, we compare the theoretical maximum
directional Shapley change obtained from the accessible-output sets with the
maximum value achieved by an explicitly recovered task. The comparison and
full verification plots are shown in \cref{fig:shift-geometry-exp}. The
largest absolute discrepancy between the two quantities is
$1.34\times10^{-10}$, with no violation among 145,000 randomly sampled tasks.
This confirms numerically that the accessible-output construction gives the
attainable maximum access-induced Shapley change for the tested finite model.

\section{Additional Results for Best-Achievable Values}
\label{app:normative-consequences}

\subsection{Corollaries of the valuation--deficiency duality}
\label{app:duality-corollaries}

\begin{corollary}[Resource simulation gap iff possible Shapley reversal]
\label{cor:resource-separation-reversal}
For fixed $\mathsf A$,
\[
\delta_{0\to1}^{\mathsf A}(\cE)>0
\]
if and only if there exists a normalized downstream task and an access-independent finite classical reference contributor that produce a strict two-contributor Shapley ranking reversal between $\cR_0$ and $\cR_1$.
\end{corollary}

\begin{corollary}[Optimal margin for a fixed task]
\label{cor:optimal-reversal-margin}
Fix $t=(\pi,u)$ and write its task-specific access advantage as
\[
\gamma_t
=
U_{\cR_1}(\pi,u;\cE)-U_{\cR_0}(\pi,u;\cE)>0.
\]
Among access-independent finite classical reference contributors, the largest possible symmetric reversal margin is exactly
\[
\frac{\gamma_t}{2}.
\]
It is attained by a classical erasure reference whose singleton value is the midpoint of the two access-dependent singleton values of contributor $B$.
\end{corollary}

The midpoint reference used above can be constructed explicitly. Let
\[
V_{\mathrm{perf}}
=
R_{\emptyset}-R_{\mathrm{perf}}^{\star},
\]
where $R_{\mathrm{perf}}^{\star}$ is the Bayes risk when the latent variable is revealed perfectly. A classically readable erasure contributor $C_s$ uses a Bernoulli reveal coin independent of $\Theta$, reveals $\theta$ with probability $s$, and returns an erasure symbol otherwise, giving
\[
v^{\star}(C_s)=sV_{\mathrm{perf}}.
\]
Choosing
\begin{equation}
s^\star
=
\frac{
v_{\cR_0}^{\star}(B)+v_{\cR_1}^{\star}(B)
}{2V_{\mathrm{perf}}}
\label{eq:optimal-erasure-s}
\end{equation}
places the reference at the midpoint and attains \cref{cor:optimal-reversal-margin}.

\begin{corollary}[Action-alphabet-free form]
\label{cor:global-duality}
Define
\[
\bar\delta_{0\to1}(\cE)
=
\sup_{\mathsf A\ \mathrm{finite}}
\delta_{0\to1}^{\mathsf A}(\cE).
\]
Then the largest symmetric Shapley ranking-reversal margin obtainable over all finite action alphabets, normalized tasks, and access-independent finite classical references is
\[
\frac12\bar\delta_{0\to1}(\cE).
\]
\end{corollary}

\begin{corollary}[Finite decision problems lift to train--test learning]
\label{cor:universal-train-test-lifting}
Any finite decision task witnessing positive deficiency or a positive valuation-reversal margin can be represented as a finite-hypothesis supervised decision-learning problem. Take a constant test input, one hypothesis $h_a$ per action $a\in\mathsf A$, set the test label to the latent variable $\Theta$, and define
\[
\ell_{\mathrm{test}}(h_a,\Theta)=L(\Theta,a).
\]
Then every subset risk, access advantage, and Shapley value is preserved.
\end{corollary}

\subsection{Corollaries of the Shapley shift geometry}
\label{app:geometry-corollaries}

For completeness, we give here the technical construction behind \cref{thm:valuation-shift-body}. For contributor $i$, expanding the change in its Shapley value gives
\begin{align}
\Delta_i^{\cR\to\cR'}(t)
=&
\sum_{S\subseteq N\setminus\{i\}}w_S
\Big(
[v_{\cR'}^{\star}(S\cup\{i\})-v_{\cR'}^{\star}(S)]\nonumber\\
&-
[v_{\cR}^{\star}(S\cup\{i\})-v_{\cR}^{\star}(S)]
\Big).
\label{eq:shapley-shift-expanded}
\end{align}
The four terms in each summand separate naturally into positive and negative parts. To combine their accessible-output sets, we use the Minkowski sum and nonnegative scalar multiplication,
\[
\mathcal X\oplus\mathcal Y:=\{x+y:x\in\mathcal X,\ y\in\mathcal Y\},
\quad
c\mathcal X:=\{cx:x\in\mathcal X\}\quad(c\ge0).
\]
Define
\begin{align}
\cK_i^+(\cR,\cR')
&=
\bigoplus_{S\subseteq N\setminus\{i\}}w_S
\left(
\mathfrak P_{\cR'}^{\mathsf A}(\cE_{S\cup\{i\}})
\oplus
\mathfrak P_{\cR}^{\mathsf A}(\cE_S)
\right),
\label{eq:cross-Ki-plus}\\
\cK_i^-(\cR,\cR')
&=
\bigoplus_{S\subseteq N\setminus\{i\}}w_S
\left(
\mathfrak P_{\cR}^{\mathsf A}(\cE_{S\cup\{i\}})
\oplus
\mathfrak P_{\cR'}^{\mathsf A}(\cE_S)
\right).
\label{eq:cross-Ki-minus}
\end{align}
For a direction $\xi\in\mathbb R^n$, write
\[
\xi_i^+=\max\{\xi_i,0\},
\quad
\xi_i^-=\max\{-\xi_i,0\},
\]
and set
\begin{align}
\cK_\xi^+(\cR,\cR')
&=
\bigoplus_i
\left(
\xi_i^+\cK_i^+(\cR,\cR')
\oplus
\xi_i^-\cK_i^-(\cR,\cR')
\right),
\label{eq:Kxi-plus}\\
\cK_\xi^-(\cR,\cR')
&=
\bigoplus_i
\left(
\xi_i^+\cK_i^-(\cR,\cR')
\oplus
\xi_i^-\cK_i^+(\cR,\cR')
\right).
\label{eq:Kxi-minus}
\end{align}
These two sets collect, respectively, the terms that increase and decrease the linear contrast selected by $\xi$.

For compact sets $\mathcal X$ and $\mathcal Y$, define the directed simulation gap
\begin{equation}
 d_{\rightarrow}(\mathcal X,\mathcal Y)
 =
 \sup_{P\in\mathcal X}
 \inf_{Q\in\mathcal Y}
 \max_{\theta\in\Theta}
 \frac12\left\|P(\cdot\mid\theta)-Q(\cdot\mid\theta)\right\|_1.
\label{eq:directed-hausdorff}
\end{equation}
The exact directional form of \cref{thm:valuation-shift-body} is
\begin{align}
\sup_{t\text{ normalized}}
\xi^\top\boldsymbol\Delta_{\cR\to\cR'}(t)
=&
\sup_{x\in\cV_{\cR\to\cR'}^{\mathsf A}}\xi^\top x \nonumber\\
=&d_{\rightarrow}\left(
 \cK_\xi^+(\cR,\cR'),
 \cK_\xi^-(\cR,\cR')
 \right).
\label{eq:directional-shift-geometry}
\end{align}

When $\cR=\cR_0\subseteq\cR_1=\cR'$, we abbreviate
\[
\cK_i^+=\cK_i^+(\cR_0,\cR_1),
\quad
\cK_i^-=\cK_i^-(\cR_0,\cR_1),
\quad
\Delta_i=\phi_i^{\cR_1,\star}-\phi_i^{\cR_0,\star}.
\]
For the fixed task, the subset access gain is
\[
g_F(S)
=
v_{\cR_1}^{\star}(S)-v_{\cR_0}^{\star}(S)
=
h_{\mathfrak P_{\cR_1}^{\mathsf A}(\cE_S)}(F)-h_{\mathfrak P_{\cR_0}^{\mathsf A}(\cE_S)}(F).
\]

\begin{proposition}[Subset access gain and Shapley redistribution]
\label{prop:premium-shapley}
For every contributor $i$,
\begin{equation}
\Delta_i(F)
=
\Sh_i(g_F)
=
h_{\cK_i^+}(F)-h_{\cK_i^-}(F),
\label{eq:delta-support-difference}
\end{equation}
and
\[
\sum_i\Delta_i(F)=g_F(N)-g_F(\varnothing).
\]
\end{proposition}

The coordinate and ranking results now follow by choosing directions in \cref{thm:valuation-shift-body}.

\begin{corollary}[Largest possible change of one contributor]
\label{thm:individual-geometry}
For $\xi=e_i$ and $\xi=-e_i$,
\begin{align}
\sup_{\pi,u}\Delta_i
&=
d_{\rightarrow}(\cK_i^+,\cK_i^-),
\label{eq:max-appreciation}\\
\sup_{\pi,u}(-\Delta_i)
&=
d_{\rightarrow}(\cK_i^-,\cK_i^+),
\label{eq:max-devaluation}\\
\sup_{\pi,u}\abs{\Delta_i}
&=
d_H(\cK_i^+,\cK_i^-).
\label{eq:max-absolute-shift}
\end{align}
In particular, contributor $i$ has identical Shapley value under the two access models for every normalized task if and only if $\cK_i^+=\cK_i^-$. Strict appreciation (respectively devaluation) occurs for some normalized task if and only if the first (respectively second) directed distance is positive.
\end{corollary}

\begin{corollary}[When an access change can affect one contributor]
\label{cor:native-invariance}
The respective criteria for invariance, possible appreciation, and possible devaluation are
\[
\cK_i^+=\cK_i^-,
\quad
 d_{\rightarrow}(\cK_i^+,\cK_i^-)>0,
\quad
 d_{\rightarrow}(\cK_i^-,\cK_i^+)>0,
\]
respectively, with the latter two interpreted as existence of a task of the corresponding sign.
\end{corollary}

For two contributors $i,j$, the direction $\xi=e_i-e_j$ gives
\begin{equation}
\cK_{ij}^+
=
\cK_i^+\oplus\cK_j^-,
\quad
\cK_{ij}^-
=
\cK_i^-\oplus\cK_j^+.
\label{eq:Kij-pm}
\end{equation}

\begin{corollary}[Largest possible pairwise ranking change]
\label{thm:pairwise-geometry}
Let
\[
G_{ij}(F)
=
\big[\phi_i^{\cR_1,\star}(F)-\phi_j^{\cR_1,\star}(F)\big]
-
\big[\phi_i^{\cR_0,\star}(F)-\phi_j^{\cR_0,\star}(F)\big].
\]
Then
\begin{align}
\sup_{\pi,u}G_{ij}
&=
d_{\rightarrow}(\cK_{ij}^+,\cK_{ij}^-),
\label{eq:max-pair-up}\\
\sup_{\pi,u}(-G_{ij})
&=
d_{\rightarrow}(\cK_{ij}^-,\cK_{ij}^+),
\label{eq:max-pair-down}\\
\sup_{\pi,u}\abs{G_{ij}}
&=
d_H(\cK_{ij}^+,\cK_{ij}^-).
\label{eq:max-pair-abs}
\end{align}
Consequently, for any fixed task with weaker-access margin
\[
m_{ij}^{(0)}
=
\phi_i^{\cR_0,\star}-\phi_j^{\cR_0,\star}>0,
\]
\begin{equation}
m_{ij}^{(0)}
>
d_{\rightarrow}(\cK_{ij}^-,\cK_{ij}^+)
\quad\Longrightarrow\quad
\phi_i^{\cR_1,\star}>\phi_j^{\cR_1,\star}.
\label{eq:geometric-no-reversal}
\end{equation}
\end{corollary}

\begin{theorem}[Distance between Shapley vectors across access models]
\label{thm:resource-valuation-pseudometric}
Let $\norm{\cdot}_{\mathsf V}$ be any norm on $\mathbb R^n$ with dual norm $\norm{\cdot}_{\mathsf V,*}$. Define
\begin{equation}
D_{\mathsf V}^{\mathsf A}(\cR,\cR')
=
\sup_{t=(\pi,u)}
\norm{
\boldsymbol\phi^{\cR,\star}(t)
-
\boldsymbol\phi^{\cR',\star}(t)
}_{\mathsf V}.
\label{eq:vector-resource-distance}
\end{equation}
Then
\begin{equation}
D_{\mathsf V}^{\mathsf A}(\cR,\cR')
=
\sup_{\norm{\xi}_{\mathsf V,*}\le1}
 d_{\rightarrow}\!\left(
 \cK_\xi^+(\cR,\cR'),
 \cK_\xi^-(\cR,\cR')
 \right).
\label{eq:vector-resource-geometry}
\end{equation}
Moreover $D_{\mathsf V}^{\mathsf A}$ is a pseudometric on admissible physical access models. It becomes a genuine metric after quotienting access models that induce the same full Shapley vector for every normalized task on $(\Theta,\mathsf A)$.

The earlier contributor-specific and pairwise-ranking discrepancies are scalar specializations:
\begin{align}
D_i^{\mathsf A}(\cR,\cR')
&=
\sup_t\abs{e_i^\top\boldsymbol\Delta_{\cR\to\cR'}(t)}\nonumber\\
&=
d_H\!\left(\cK_i^+(\cR,\cR'),\cK_i^-(\cR,\cR')\right),
\label{eq:resource-valuation-distance}\\
D_{ij}^{\mathsf A}(\cR,\cR')
&=
\sup_t\abs{(e_i-e_j)^\top\boldsymbol\Delta_{\cR\to\cR'}(t)},
\label{eq:pair-resource-pseudometric}
\end{align}
and each is itself a pseudometric (possibly more degenerate than the full-vector one).
\end{theorem}

The closed convex set formulation is the most informative object: individual appreciation/devaluation, pairwise ranking drift, and norm-based access-model distances are all projections or support-function summaries of the same closed convex envelope $\cV_{\cR\to\cR'}^{\mathsf A}$.

The following results are consequences or sanity checks of the main accessible-output formulation. They are moved out of the main narrative because they are not needed to state the exact Shapley shift geometry.

\subsection{Classically readable data remove the access gap}
\begin{theorem}[Classically readable data give the same value]
\label{thm:classical-collapse}
Fix a coalition $S$ and nested physical access models $\cR_0\subseteq\cR_1$. Suppose that the family $\{\rho_{\theta,S}\}_{\theta}$ is diagonal in a common orthonormal basis $\{\ket z\}$ and that $\cR_0$ can measure this basis and apply arbitrary classical post-processing. Then for every bounded decision problem whose only quantum input is $\rho_{\theta,S}$,
\[
R_{\cR_0}^{\star}(S)=R_{\cR_1}^{\star}(S),
\quad
v_{\cR_0}^{\star}(S)=v_{\cR_1}^{\star}(S).
\]
If the comparison is LOCC versus global measurement, the same conclusion holds when the common readable basis is a product basis available to the LOCC protocol.
\end{theorem}

\begin{remark}
Commutativity alone does not imply LOCC implementability of an arbitrary global eigenbasis. The product-basis clause is necessary when the weaker access model is local.
\end{remark}

\subsection{Resource-simulation-gap bounds}

The exact Hausdorff quantities retain all subset accessible-output sets and may themselves be difficult to evaluate. The resource simulation gap gives a simpler bound. For each coalition write
\[
\delta(S)
=
\delta_{0\to1}^{\mathsf A}(\cE_S).
\]
Because every fixed-task premium obeys
\[
0\le g_F(S)\le\delta(S),
\]
the exact geometry immediately yields the following bounds.

\begin{corollary}[Resource-simulation-gap bounds on individual Shapley shifts]
\label{thm:deficiency-premium}
For every contributor $i$,
\begin{align}
d_{\rightarrow}(\cK_i^+,\cK_i^-)
&\le
\bbE_{S\sim\Sh_i}\delta(S\cup\{i\}),
\label{eq:deficiency-appreciation}\\
d_{\rightarrow}(\cK_i^-,\cK_i^+)
&\le
\bbE_{S\sim\Sh_i}\delta(S).
\label{eq:deficiency-devaluation}
\end{align}
Hence every fixed task satisfies
\begin{equation}
-\bbE_{S\sim\Sh_i}\delta(S)
\le
\Delta_i
\le
\bbE_{S\sim\Sh_i}\delta(S\cup\{i\}).
\label{eq:deficiency-delta}
\end{equation}
In particular, with
\[
\delta_\star^{\mathsf A}
=
\max_{T\subseteq N}\delta(T),
\]
\[
\abs{\Delta_i}\le\delta_\star^{\mathsf A}.
\]
\end{corollary}

\begin{corollary}[Tractable no-reversal certificates]
\label{cor:no-reversal}
For a fixed task with weaker-access margin $m_{ij}^{(0)}>0$,
\begin{equation}
m_{ij}^{(0)}
>
\bbE_{S\sim\Sh_i}\delta(S)
+
\bbE_{T\sim\Sh_j}\delta(T\cup\{j\})
\Longrightarrow
\phi_i^{\cR_1,\star}>\phi_j^{\cR_1,\star}.
\label{eq:weighted-no-reversal}
\end{equation}
The simpler sufficient condition
\[
m_{ij}^{(0)}>2\delta_\star^{\mathsf A}
\]
follows immediately. 
\end{corollary}

\subsection{Distance to access equivalence}
Let
\[
\cC_{0,1}^{\mathsf A}
=
\left\{
\cF=\{\sigma_\theta\}:
\delta_{0\to1}^{\mathsf A}(\cF)=0\right\},
\]
where $\cF$ is defined on the same Hilbert space and access-model realization as $\cE$. 
This is an access-equivalence (or access-insensitivity) set, not a classical-state set: its members need not commute or be classically readable; they are simply experiments for which the two access models have zero resource simulation gap on the specified action class. The comparison set is nonempty under Assumption~\ref{ass:shared-baseline}: any $\theta$-independent experiment on the same Hilbert space has identical accessible classical output behaviors under the two access models, because its output cannot carry information about $\theta$ and both access models have the same free classical randomization and post-processing.
Define
\begin{equation}
q_{0,1}^{\mathsf A}(\cE)
=
\inf_{\cF=\{\sigma_\theta\}\in\cC_{0,1}^{\mathsf A}}
\sup_\theta
\frac12\norm{\rho_\theta-\sigma_\theta}_1.
\label{eq:q-resource-equivalence}
\end{equation}

\begin{theorem}[Distance to access equivalence controls access dependence]
\label{thm:q-bound}
For every experiment $\cE$,
\begin{equation}
\delta_{0\to1}^{\mathsf A}(\cE)
\le
2q_{0,1}^{\mathsf A}(\cE).
\label{eq:q-to-deficiency}
\end{equation}
Hence, with
\[
q_\star^{\mathsf A}
=
\max_{T\subseteq N}q_{0,1}^{\mathsf A}(\cE_T),
\]
\[
\abs{\phi_i^{\cR_1,\star}-\phi_i^{\cR_0,\star}}
\le2q_\star^{\mathsf A}.
\]
An initial valuation margin greater than $4q_\star^{\mathsf A}$ therefore cannot reverse.
\end{theorem}

The best-achievable analysis now has two exact levels. The resource-simulation-gap theorem characterizes the largest benchmarked reversal that can witness a difference between access models. Shapley shift geometry characterizes the largest appreciation, devaluation, and pairwise ranking drift of the contributors already present in an arbitrary game. The resource simulation gap and distance to access equivalence are simpler summaries of this native geometry and provide task-uniform certificates when the exact aggregate sets are difficult to evaluate.

\section{Secondary Theoretical Consequences}
\label{app:secondary-consequences}

\subsection{From a fixed learner to the best-achievable value}
\label{app:normative-operational-bridge}
The fixed-learner value is exact for a specified learner. When that learner uniformly approximates the access-model optimum, the same measured attribution also approximates the best-achievable attribution.

\begin{theorem}[Fixed-learner approximation to the best-achievable value]
\label{thm:normative-operational-bridge}
Let a fixed learner be feasible in $\cR$ and use the same no-data optimum. If, for every coalition,
\begin{equation}
0\le v_{\cR}^{\star}(S)-v_{\cA}(S)\le\eta,
\label{eq:eta-suboptimality}
\end{equation}
then
\begin{equation}
|\phi_i^{\cR,\star}-\phi_i^{\cA}|\le\eta\quad\forall i.
\label{eq:bridge-shapley}
\end{equation}
If $\widehat\phi_i^{\cA}$ has estimation error at most $\epsilon$, then its total error relative to the best-achievable value is at most $\eta+\epsilon$.
\end{theorem}

Combining this bridge with \cref{thm:valuation-deficiency-duality} turns a sufficiently resolved observed ranking flip into a resource certificate.
\begin{corollary}[Finite-copy resource certificate]
\label{cor:finite-copy-resource-certificate}
In the two-contributor benchmark setting, if all four estimated values are within $\tau=\eta+\epsilon$ of their best-achievable counterparts and $\widehat m$ is the observed symmetric reversal margin, then
\begin{equation}
\delta_{0\to1}^{\mathsf A}(\cE_B)\ge 2[\widehat m-2\tau]_+,
\label{eq:finite-copy-deficiency-certificate}
\end{equation}
and
\begin{equation}
q_{0,1}^{\mathsf A}(\cE_B)\ge[\widehat m-2\tau]_+.
\label{eq:finite-copy-resource-insensitivity-certificate}
\end{equation}
\end{corollary}

\subsection{Single-contributor copy complexity}

Let
\[
W_i:=\lambda_{\max}(\Omega_i)-\lambda_{\min}(\Omega_i)\le2
\]
be the spectral width of the Shapley observable.

\begin{proposition}[Single-valuation copy complexity]
\label{prop:single-copy-complexity}
For known implementable $\Omega_i$ of width $W_i>0$, an $\epsilon$-accurate estimate with failure probability $p_{\mathrm{fail}}$ is achievable with
\begin{equation}
K=O\!\left(
\frac{W_i^2}{\epsilon^2}
\log\frac1{p_{\mathrm{fail}}}
\right).
\label{eq:single-copy-upper-main}
\end{equation}
Conversely, for every $0<W\le1$, $0<\epsilon\le W/8$, and $0<p_{\mathrm{fail}}\le1/4$, there is a one-qubit fixed-learner instance of width $W$ requiring
\begin{equation}
K=\Omega\!\left(
\frac{W^2}{\epsilon^2}
\log\frac1{p_{\mathrm{fail}}}
\right)
\label{eq:single-copy-lower-main}
\end{equation}
even under collective measurements across the $K$ replicas.
\end{proposition}

Here one replica means one fresh preparation of the complete finite-copy training block. In an independent-source implementation, the physical preparation cost for each contributor is therefore multiplied by the number of replicas used by the estimator.

\subsection{Simultaneous fixed-learner valuation and shadows}
\label{app:shadow-reduction}
The family
\[
\{\Omega_1,\ldots,\Omega_n\}
\]
turns simultaneous data valuation into a many-observable estimation problem.

\begin{theorem}[Shadow reduction for fixed-learner valuation]
\label{thm:shadow-reduction}
Any measurement scheme that, from $K$ replicas of an unknown state, simultaneously estimates the expectations of a known observable family $\{O_j\}_{j=1}^{J}$ to accuracy $\epsilon$ and failure probability $p_{\mathrm{fail}}$ can be applied directly to \raqdv\ by setting $O_j=\Omega_j$. In particular, under a classical-shadow measurement ensemble for which
\[
\max_i\norm{\Omega_i}_{\mathrm{sh}}^2\le\Xi_{\mathrm{sh}},
\]
the standard shadow-tomography and classical-shadow guarantees \citep{aaronson2018shadow,huang2020shadows} imply
\[
K
=
O\!\left(
\frac{\Xi_{\mathrm{sh}}}{\epsilon^2}
\log\frac{n}{p_{\mathrm{fail}}}
\right)
\]
replicas for simultaneous estimation of all $n$ fixed-learner valuations.
\end{theorem}

This is a reduction, not a new shadow-tomography theorem. Its significance is physical: the same destroyed copies can support many valuation estimates rather than requiring fresh quantum training data for every contributor. Computational efficiency is a separate issue because explicitly forming $\Omega_i$ may require exponentially many subset effects.

\begin{remark}[Permutation implementation]
If $\Omega_i$ cannot be explicitly assembled, one may sample Shapley permutations and estimate only the corresponding prefix-subset effects. This avoids enumerating all $2^{n-1}$ coalitions but still requires an implementation of the sampled effects. The resulting method addresses physical copy reuse; it does not by itself solve the computational complexity of training or effect synthesis.
\end{remark}

This appendix collects consequences that are mathematically useful but secondary to the paper's main access-relative valuation geometry.

\subsection{Information-valued Shapley value function}
\label{sec:info}

As a complementary information-valued cooperative game, outside the bounded-loss normalization used in the decision-theoretic results above, consider a classical target with quantum side information. The following decomposition is a direct consequence of the quantum mutual-information chain rule and is included to separate intrinsic information content from access-constrained accessibility.

Let
\[
\omega_{YQ_1\cdots Q_n}
=
\sum_y p_y\ketbra{y}{y}\otimes\rho_y^{Q_1\cdots Q_n}
\]
be a classical--quantum state. Define the intrinsic coalition information
\begin{equation}
v_\chi(S)=I(Y:Q_S)_\omega.
\label{eq:holevo-game}
\end{equation}
Because $Y$ is classical, this is the Holevo information of the corresponding ensemble \citep{holevo1973bounds}.

\begin{proposition}[Information-Shapley decomposition]
\label{prop:info-shapley}
Define $\psi_i^\chi:=\Sh_i(v_\chi)$. Then
\begin{equation}
\psi_i^\chi
=
\bbE_{S\sim\Sh_i}
I(Y:Q_i\mid Q_S)_\omega.
\label{eq:cmi-shapley}
\end{equation}
Hence $\psi_i^\chi\ge0$ and
\[
\sum_i\psi_i^\chi
=
I(Y:Q_1\cdots Q_n)_\omega.
\]
\end{proposition}

Now let $\cR$ be a measurement physical access model and define access-constrained accessible information
\[
A_{\cR}(S)
=
\sup_{M\in\cR}I(Y:Z_M),
\quad
\phi_i^{\cR,\mathrm{info}}:=\Sh_i(A_{\cR}).
\]
Holevo's bound gives $A_{\cR}(S)\le v_\chi(S)$. Define the accessibility deficit
\[
D_{\cR}(S)=v_\chi(S)-A_{\cR}(S)\ge0.
\]
Then the access-constrained information-game valuation decomposes as
\begin{equation}
\phi_i^{\cR,\mathrm{info}}
=
\psi_i^\chi-\Sh_i(D_{\cR}).
\label{eq:intrinsic-operational}
\end{equation}
For nested access models,
\[
\phi_i^{\cR_1,\mathrm{info}}-\phi_i^{\cR_0,\mathrm{info}}
=
\Sh_i(D_{\cR_0}-D_{\cR_1}).
\]
Thus the intrinsic quantum information of the data does not change under an access upgrade; what changes is the fraction of that information that the measurement model can operationally unlock.

\subsection{Valuation degree inherited from multi-copy learning hierarchies}
\label{sec:degree}

The best-achievable analysis also gives a simple way to summarize how much coherent copy access is required before a dataset becomes valuable.

\begin{definition}[Budgeted valuation degree]
\label{def:valuation-degree}
Let $\mathbf D=\{D_m\}_{m\ge1}$ be a family of train--test instances indexed by problem size $m$, with full contributor set $N_m$ in instance $D_m$. Fix nested coherent-access physical access models
\[
\cR^{(1)}\subseteq\cR^{(2)}\subseteq\cdots
\]
and a total state-preparation/sample budget $\mathsf B(m)$. We write $\cR^{(k)}[\mathsf B(m)]$ for the protocols in $\cR^{(k)}$ that obey this budget; for the intended copy hierarchy, $\cR^{(k)}$ may be read as protocols whose coherent quantum measurement blocks act on at most $k$ training copies at a time, with the classical processing allowed by the access model. Define the budget-restricted full-dataset risk and value by
\begin{align}
R_{\cR^{(k)}}^{\star,\mathsf B}(D_m)
:=&
\inf_{\cA\in\cR^{(k)}[\mathsf B(m)]}
R_{\mathrm{test}}(\cA;D_m),\\
v_{\cR^{(k)}}^{\star,\mathsf B}(D_m)
:=&
R_{\emptyset,m}-R_{\cR^{(k)}}^{\star,\mathsf B}(D_m).
\label{eq:budgeted-dataset-value}
\end{align}
For a fixed target value $\tau>0$, the asymptotic valuation degree is
\begin{equation}
d_{\tau,\mathsf B}(\mathbf D)
:=
\min\left\{
k\ge1:
\exists m_0\ \forall m\ge m_0,\quad
v_{\cR^{(k)}}^{\star,\mathsf B}(D_m)\ge\tau
\right\},
\label{eq:valuation-degree}
\end{equation}
with $d_{\tau,\mathsf B}(\mathbf D)=\infty$ if the set is empty.
\end{definition}

\begin{proposition}[Classical degree collapse]
\label{prop:degree-classical}
Suppose that, for every $m$, the full supplied training state in $D_m$ is diagonal in a tensor-product basis obtained from a common readable single-copy basis, and that the access model permits measuring that basis independently on every preparation followed by arbitrary classical processing. If $d_{\tau,\mathsf B}(\mathbf D)<\infty$, then
\[
d_{\tau,\mathsf B}(\mathbf D)=1.
\]
\end{proposition}

\citet{noller2025hierarchy} construct quantum learning tasks for which $k$-copy measurements are efficient while every smaller-copy strategy is exponentially sample-inefficient, for infinitely many $k$. Their hierarchy transfers to valuation once its learning criterion is put on a bounded train--test utility scale. We state the transfer explicitly because the valuation result is a corollary of their sample-complexity separation rather than a new multi-copy lower bound.

\begin{corollary}[Unbounded quantum valuation degree]
\label{cor:degree-hierarchy}
Fix a copy degree $k$ for which the hierarchy of \citet{noller2025hierarchy} supplies a constant-gap learning separation: under a common bounded utility encoding there are constants $u_{\rm hi}>u_{\rm lo}$ such that a $k$-copy learner reaches utility at least $u_{\rm hi}$ with polynomially many preparations, whereas every smaller-copy learner using any polynomial number of preparations has utility at most $u_{\rm lo}$ for all sufficiently large instances. If the bounded utility encoding has a common no-data optimum $u_\emptyset\le u_{\rm lo}$ across the family, then there exist a quantum train--test family $\mathbf D$, a polynomial budget $\mathsf B(m)$, and a fixed threshold $\tau>0$ such that
\[
d_{\tau,\mathsf B}(\mathbf D)=k.
\]
Thus any constant-gap copy-complexity hierarchy induces the corresponding valuation-degree hierarchy. This is a transfer of the learning separation, not a new copy-complexity lower bound.
\end{corollary}

\subsection{Valuation, disturbance, and locally gentle estimation}
\label{sec:disturbance}

Quantum valuation has a second physical cost: the act of learning a value may disturb the state that is being valued. The correct statement requires care.

\subsubsection{No universal loss for a predetermined future task}

\begin{proposition}[No universal fixed-task damage]
\label{prop:no-universal-task}
There is no strictly positive function $f_{\mathrm{dmg}}$ such that every nontrivial valuation protocol satisfying a prescribed accuracy must reduce the performance of every predetermined future learning task by at least $f_{\mathrm{dmg}}$.
\end{proposition}

Indeed, take
\[
\rho_\theta=\tau_\theta^V\otimes\sigma_\theta^T
\]
and let valuation measure only subsystem $V$, while the future task depends only on $T$. Valuation can be arbitrarily informative about a functional of $\tau_\theta^V$ while leaving $\sigma_\theta^T$ unchanged. Any valid universal tradeoff must therefore be phrased in terms of recoverability, worst-case future tasks, or an explicit coupling between valuation-relevant and task-relevant degrees of freedom.

\subsubsection{Accurate valuation leaks information}

Consider a valuation instrument
\[
\mathcal I_{\mathrm{val}}:Q\to ZQ_{\mathrm{res}},
\]
where $Z$ is a classical valuation transcript and $Q_{\mathrm{res}}$ is the residual quantum system. Let
\[
\cN=\Tr_Z\circ\mathcal I_{\mathrm{val}}
\]
be the residual channel.

\begin{lemma}[Valuation accuracy implies transcript distinguishability]
\label{lem:valuation-distinguishability}
Let $\rho_0,\rho_1$ be two possible training states with
\[
|\phi(\rho_0)-\phi(\rho_1)|>2\epsilon.
\]
If an estimator computed from $Z$ is $\epsilon$-accurate on each state with probability at least $1-p_{\mathrm{fail}}$, then
\[
\TV(P_{Z|0},P_{Z|1})\ge1-2p_{\mathrm{fail}}.
\]
\end{lemma}

The proof is the midpoint test: an $\epsilon$-accurate estimate identifies which side of the midpoint contains the true valuation.

\subsubsection{Residual-only no-free valuation}

Define the residual-only recoverability error
\begin{equation}
\eta_{\mathrm{rec}}(\cN)
=
\inf_{\cR_{\mathrm{rec}}}
\norm{
\cR_{\mathrm{rec}}\circ\cN-\id_Q
}_{\diamond},
\label{eq:recoverability-error}
\end{equation}
where the recovery channel can use only $Q_{\mathrm{res}}$, not the classical valuation transcript $Z$.

\begin{theorem}[Residual-only no-free valuation]
\label{thm:residual-no-free}
Under the standard unnormalized diamond-norm convention, if the assumptions of \cref{lem:valuation-distinguishability} hold, then the residual channel cannot be uniformly close to perfectly recoverable. In particular, the information--disturbance theorem of \citet{ksw2006} implies a universal lower bound of the form
\[
\eta_{\mathrm{rec}}(\cN)
\ge
\frac{(1-2p_{\mathrm{fail}})^2}{4}.
\]
Consequently, by the ancilla-assisted randomization/deficiency characterization of quantum experiments, there exists a downstream quantum decision problem within that comparison class for which using only the residual system is strictly worse than using the untouched input.
\end{theorem}

The theorem does \emph{not} claim that every fixed downstream task loses performance. It says that reliable valuation is incompatible with retaining a residual channel that is uniformly as informative as the identity for all future decisions.

\subsubsection{Finite-copy gentle valuation}

We now obtain a confidence-sensitive finite-copy tradeoff in the locally gentle model of \citet{butucea2025gentle}, building on the broader gentle-measurement perspective of \citet{aaronson2019gentle}. Let
\[
\phi_\Omega(\rho)=\Tr(\Omega\rho)
\]
with spectral width
\[
W_\Omega
=
\lambda_{\max}(\Omega)-\lambda_{\min}(\Omega)>0.
\]

We use the trace-distance convention
\[
T(\rho,\sigma)=\frac12\norm{\rho-\sigma}_1.
\]
The strong qDPI of \citet{butucea2025gentle} states that for an $\alpha$-gentle measurement with $\alpha<1/2$,
\begin{equation}
D_{\mathrm{KL}}^{\mathrm{sym}}
(P_{\rho},P_{\sigma})
\le
\left(
\frac{8\alpha}{(1-2\alpha)^2}
\right)^2
T(\rho,\sigma)^2.
\label{eq:gentle-qdpi}
\end{equation}
Following the terminology of \citet{butucea2025gentle}, gentleness is always relative to a specified state class. In this paper we use the strongest state-independent minimax convention: every single-copy factor must be $\alpha$-gentle on the full state space $\cS(\cH)$ of the system on which $\Omega$ acts. A non-adaptive product protocol
\[
M^{(K)}=\bigotimes_{\ell=1}^K M_\ell
\]
is called \emph{locally $\alpha$-gentle} when each $M_\ell$ belongs to this full-state-space gentle class. This convention contains every hard state used below and makes the supremum over $\rho$ in the minimax definition unambiguous. For independent copies and non-adaptive product measurements, the symmetrized KL divergence adds across copies.

\begin{definition}[Locally gentle valuation complexity]
\label{def:gentle-val-complexity}
For a known Hermitian observable $\Omega$ with $W_\Omega>0$, define $C_{\mathrm{gentle\text{-}val}}(\Omega,\epsilon,p_{\mathrm{fail}};\alpha)$ as the smallest integer $K$ for which there exist a non-adaptive product protocol $M^{(K)}=\bigotimes_{\ell=1}^K M_\ell$ whose every factor is $\alpha$-gentle on the full state space $\cS(\cH)$, and a classical estimator $\widehat\phi$ satisfying
\begin{equation}
\sup_{\rho}
\bbP_{\rho^{\otimes K}}
\left(
\abs{\widehat\phi-\Tr(\Omega\rho)}>\epsilon
\right)
\le p_{\mathrm{fail}}.
\label{eq:gentle-val-complexity-def}
\end{equation}
Here the supremum ranges over all density operators on the Hilbert space on which $\Omega$ acts. The protocol and estimator may depend on the known observable $\Omega$, but not on the unknown state $\rho$; if no such finite $K$ exists, the complexity is defined to be $\infty$.
\end{definition}

\begin{theorem}[Confidence-tight locally gentle valuation lower bound]
\label{thm:gentle-lower}
Assume $0<\epsilon\le W_\Omega/8$, $0<p_{\mathrm{fail}}<1/2$, and $K$ independent copies are measured non-adaptively by a locally $\alpha$-gentle product protocol $\bigotimes_{\ell=1}^K M_\ell$, with $0<\alpha<1/2$. If the resulting estimator is uniformly $\epsilon$-accurate with failure probability at most $p_{\mathrm{fail}}$,
\[
\bbP_\rho\big(
|\widehat\phi-\phi_\Omega(\rho)|\le\epsilon
\big)
\ge1-p_{\mathrm{fail}}
\quad
\text{for every }\rho,
\]
then
\begin{equation}
K
\ge
\frac{
W_\Omega^2(1-2\alpha)^4
}{
512\,\alpha^2\epsilon^2
}
(1-2p_{\mathrm{fail}})
\log\frac{1-p_{\mathrm{fail}}}{p_{\mathrm{fail}}}.
\label{eq:gentle-lower-bound}
\end{equation}
In particular, for every fixed $\alpha_0<1/2$, uniformly over $0<\alpha\le\alpha_0$ and $0<p_{\mathrm{fail}}\le1/4$,
\begin{equation}
K
=
\Omega_{\alpha_0}\!\left(
\frac{W_\Omega^2}{\alpha^2\epsilon^2}
\log\frac{1}{p_{\mathrm{fail}}}
\right).
\label{eq:gentle-confidence-lower}
\end{equation}
\end{theorem}

\begin{corollary}[Quantum Label Switch achievability]
\label{cor:gentle-upper}
Normalize the Shapley observable to the two-outcome effect
\[
E
=
\frac{\Omega-\lambda_{\min}\1}{W_\Omega},
\quad
0\le E\le\1.
\]
Realize the binary POVM $\{E,\1-E\}$ by a Naimark dilation: append a fixed ancilla, apply a unitary $V_{\mathrm N}$, and measure a binary projector $\Pi$ on the enlarged system so that the projective outcome probability equals $\Tr(E\rho)$. Apply the quantum Label Switch construction of \citet{butucea2025gentle} to this binary projective measurement on the full enlarged state space. After recording its classical binary output, apply $V_{\mathrm N}^\dagger$ and discard the ancilla. Unitary invariance and trace-distance contractivity under the final partial trace imply that the induced instrument on the original system is $\alpha$-gentle for every input state. Applying this construction independently across replicas therefore yields an admissible locally $\alpha$-gentle product protocol. The resulting binary observation has signal attenuation
\begin{equation}
\beta_\alpha
=
\frac{2\alpha}{1+\alpha^2}.
\label{eq:qls-attenuation}
\end{equation}
After undoing this known attenuation, Hoeffding concentration shows that
\begin{equation}
K
\ge
\frac{W_\Omega^2}{2\beta_\alpha^2\epsilon^2}
\log\frac{2}{p_{\mathrm{fail}}}
\label{eq:gentle-upper-bound}
\end{equation}
suffices for additive valuation error $\epsilon$ with failure probability at most $p_{\mathrm{fail}}$.
Consequently, for every fixed $\alpha_0<1/2$, $0<\alpha\le\alpha_0$, $0<p_{\mathrm{fail}}\le1/4$, and $0<\epsilon\le W_\Omega/8$,
\begin{equation}
C_{\mathrm{gentle\text{-}val}}
(\Omega,\epsilon,p_{\mathrm{fail}};\alpha)
=
\Theta_{\alpha_0}\!\left(
\frac{W_\Omega^2}{\alpha^2\epsilon^2}
\log\frac{1}{p_{\mathrm{fail}}}
\right).
\label{eq:gentle-minimax-tight}
\end{equation}
Thus the locally gentle valuation problem is tight not only in $(W_\Omega,\alpha,\epsilon)$ but also in its confidence dependence.
\end{corollary}

For a fixed-learner Shapley value, \cref{thm:valuation-observable} gives $W_{\Omega_i}\le2$. Hence contributor valuation inherits the same confidence-tight locally gentle copy-complexity law.

\section{Notation for Best-Achievable and Fixed-Learner Values}
\label{app:notation-scope}
\label{app:notation}

This appendix records the information contract used by every theorem.

\paragraph{Best-achievable value.}
For each coalition $S$, $v_{\cR}^{\star}(S)$ is defined by optimizing over the physical access model $\cR$. The optimizer may depend on $S$. The compared access models share the no-data baseline, allow free disposal of supplied quantum systems, and allow free classical processing. For the valuation--deficiency theorem we fix finite $\Theta$ and $\mathsf A$ and work with the closed convex accessible-output sets $\mathfrak P_{\cR_r}^{\mathsf A}(\cE)$ from \cref{eq:interface-set}.

\paragraph{Fixed-learner value.}
A single learning rule is fixed before valuation; $\cA_S$ denotes its coalition-$S$ instantiation. Randomness internal to the learner and randomness in the independent test draw are averaged into the expected test risk $R_{\cA}(S)$. The fixed-learner game is $v_{\cA}(S)=R_{\cA}(\varnothing)-R_{\cA}(S)$. When comparing it directly with the best-achievable game, we additionally require the fixed learner's no-data rule to attain the shared optimum, so that $R_{\cA}(\varnothing)=R_{\emptyset}$.

\paragraph{Why the distinction is mathematically necessary.}
A fixed quantum channel/instrument followed by a bounded score defines a linear functional of the input state. An optimum over a family of channels is a pointwise supremum of such functionals and is generally convex rather than linear. The observable representation therefore cannot be transferred from $v_{\cA}$ to an unrestricted $v_{\cR}^{\star}$ without an additional finite-class, covering-number, or duality argument.

\paragraph{Index and accuracy conventions.}
We reserve $n$ for the number of data contributors, $c_i$ for the physical copy multiplicity supplied by contributor $i$ inside one training block, $K$ for the number of fresh replicas used by an estimation protocol (replicas of the complete training block in contributor-valuation applications), and $k$ for coherent copy degree in the multi-copy hierarchy. The symbol $\delta_{0\to1}^{\mathsf A}$ denotes resource simulation gap (one-sided deficiency), while $p_{\mathrm{fail}}$ denotes an estimation failure probability. Keeping these symbols separate avoids overloading $\delta$ with two unrelated meanings.

\paragraph{Distance conventions.}
For states, $T(\rho,\sigma)=\frac12\norm{\rho-\sigma}_1$. This is the quantity denoted $\norm{\rho-\sigma}_{\mathrm{Tr}}$ by \citet{butucea2025gentle}. For classical distributions, $\TV(P,Q)=\frac12\norm{P-Q}_1$. The residual recoverability theorem uses the standard unnormalized diamond norm for channels. The confidence-tight gentle lower bound below is written directly in this trace-distance convention.

\section{Proofs for Best-Achievable Values}

\subsection{Proof of \cref{thm:classical-collapse}}
\label{app:proof-classical-collapse}
Write
\[
\rho_{\theta,S}
=
\sum_z p_\theta(z)\ketbra z z.
\]
Let $M=\{M_a\}_a$ be any POVM allowed by the richer access model. Its outcome law is
\[
P_\theta^M(a)
=
\Tr(M_a\rho_{\theta,S})
=
\sum_z p_\theta(z)\bra z M_a\ket z.
\]
Define $\kappa(a|z)=\bra z M_a\ket z$. Positivity gives $\kappa(a|z)\ge0$, and $\sum_aM_a=\1$ gives $\sum_a\kappa(a|z)=1$. Thus $\kappa$ is a classical stochastic kernel. A protocol that first measures the common basis and then samples $a\sim\kappa(\cdot|z)$ exactly reproduces the stronger access model's outcome distribution for every $\theta$. By assumption this basis measurement and arbitrary classical post-processing are available to $\cR_0$. Hence every decision rule implementable under $\cR_1$ is statistically simulable under $\cR_0$. Since $\cR_0\subseteq\cR_1$, the optimal risks are equal. The LOCC variant follows identically when the common basis is a product basis locally measurable by the weaker access model. \qed

\subsection{Proof of \cref{thm:valuation-deficiency-duality}}
\label{app:proof-valuation-deficiency}
We separate the proof into the accessible-output duality and the Shapley characterization.

\paragraph{Step 1: the relevant output-behavior norm.}
For two conditional action arrays $P,Q$, set $X=P-Q$. Every row has zero sum:
\[
\sum_aX_{\theta a}=0.
\]
Define
\[
\norm{X}_{\infty,\TV}
=
\max_\theta\frac12\sum_a|X_{\theta a}|.
\]
For a real array $F=(F_{\theta a})$, let
\[
\operatorname{osc}(F_\theta)
=
\max_aF_{\theta a}-\min_aF_{\theta a}
\]
and
\[
\mathfrak F_\star
=
\left\{
F:
\sum_\theta\operatorname{osc}(F_\theta)\le1
\right\}.
\]
Then
\begin{equation}
\norm{X}_{\infty,\TV}
=
\sup_{F\in\mathfrak F_\star}
\sum_{\theta,a}F_{\theta a}X_{\theta a}.
\label{eq:dual-interface-norm}
\end{equation}
To see the upper bound, shift each row of $F$ by $-\min_aF_{\theta a}$. This leaves the pairing unchanged because each row of $X$ sums to zero. The shifted row lies in $[0,r_\theta]$ with $r_\theta=\operatorname{osc}(F_\theta)$. Hence
\[
\sum_aF_{\theta a}X_{\theta a}
\le
r_\theta\frac12\norm{X_\theta}_1,
\]
and summing over $\theta$ gives at most $\norm{X}_{\infty,\TV}$ because $\sum_\theta r_\theta\le1$. For achievability, choose a row $\theta^\star$ attaining the maximum and set $F_{\theta^\star a}=\mathbf 1\{X_{\theta^\star a}>0\}$, with all other rows zero. The pairing is exactly the positive mass of that zero-sum row, namely $\frac12\norm{X_{\theta^\star}}_1$.

Because row constants never affect a difference of conditional distributions, we may henceforth use the normalized representative set
\begin{equation}
\mathfrak F
=
\left\{
F_{\theta a}\ge0:
\sum_\theta\max_aF_{\theta a}\le1
\right\}.
\label{eq:normalized-dual-set}
\end{equation}

\paragraph{Step 2: distance to the weaker accessible-output set.}
Fix $P\in\mathfrak P_{\cR_1}^{\mathsf A}(\cE)$. Since $\mathfrak P_{\cR_0}^{\mathsf A}(\cE)$ and $\mathfrak F$ are convex compact sets in finite-dimensional spaces and the pairing is bilinear and continuous, the standard minimax theorem gives
\begin{align}
\inf_{Q\in\mathfrak P_{\cR_0}^{\mathsf A}}
\norm{P-Q}_{\infty,\TV}
&=
\inf_Q\sup_{F\in\mathfrak F}\langle F,P-Q\rangle\nonumber\\
&=
\sup_{F\in\mathfrak F}
\left[
\langle F,P\rangle
-
\sup_{Q\in\mathfrak P_{\cR_0}^{\mathsf A}}\langle F,Q\rangle
\right].
\label{eq:minimax-distance}
\end{align}
Taking the supremum over $P\in\mathfrak P_{\cR_1}^{\mathsf A}$ yields
\begin{equation}
\delta_{0\to1}^{\mathsf A}(\cE)
=
\sup_{F\in\mathfrak F}
\left[
\sup_{P\in\mathfrak P_{\cR_1}^{\mathsf A}}\langle F,P\rangle
-
\sup_{Q\in\mathfrak P_{\cR_0}^{\mathsf A}}\langle F,Q\rangle
\right].
\label{eq:deficiency-support-gap}
\end{equation}

\paragraph{Step 3: every normalized dual functional is prior times payoff.}
Let $F\in\mathfrak F$ and set $r_\theta=\max_aF_{\theta a}$. Because $\sum_\theta r_\theta\le1$, choose a prior $\pi$ with $\pi_\theta\ge r_\theta$ and $\sum_\theta\pi_\theta=1$. Define
\[
u(\theta,a)=
\begin{cases}
F_{\theta a}/\pi_\theta,&\pi_\theta>0,\\
0,&\pi_\theta=0.
\end{cases}
\]
Then $0\le u\le1$ and $F_{\theta a}=\pi_\theta u(\theta,a)$. Conversely every pair $(\pi,u)$ with $0\le u\le1$ defines an element $F=\pi u$ of $\mathfrak F$, because
\[
\sum_\theta\max_aF_{\theta a}
\le
\sum_\theta\pi_\theta=1.
\]
Substitution into \cref{eq:deficiency-support-gap} gives
\[
\delta_{0\to1}^{\mathsf A}(\cE)
=
\sup_{\pi,u}
\big[
U_{\cR_1}(\pi,u;\cE)-U_{\cR_0}(\pi,u;\cE)
\big].
\]
This is the finite-output decision-theoretic half of the theorem.

\paragraph{Step 4: from task advantage to the optimal Shapley reversal.}
Fix a task $t=(\pi,u)$ and write its access advantage as
\[
\gamma_t
=
U_{\cR_1}(\pi,u;\cE)-U_{\cR_0}(\pi,u;\cE)
\ge0.
\]
Let $v_0$ and $v_1$ be contributor $B$'s singleton values under the two access models. Baseline subtraction cancels, so $v_1-v_0=\gamma_t$. If $\gamma_t=0$, no positive symmetric reversal is possible against an access-independent reference. Suppose $\gamma_t>0$. Perfect knowledge of $\Theta$ cannot be worse than either access model, so the perfect-information value $V_{\mathrm{perf}}$ satisfies
\[
0\le v_0<v_1\le V_{\mathrm{perf}}.
\]
Construct a classically readable erasure reference $C_s$ using a reveal coin independent of $\Theta$; it reveals $\Theta$ with probability $s$ and outputs an erasure symbol otherwise. By Assumption~\ref{ass:shared-baseline},
\[
v(C_s)=sV_{\mathrm{perf}}
\]
under both access models. Choose
\[
s^\star=\frac{v_0+v_1}{2V_{\mathrm{perf}}},
\]
which lies in $(0,1]$. Let the joint two-contributor experiment be conditionally independent given $\theta$,
\[
\rho_{\theta,CB}=\rho_{\theta,C_s}\otimes\rho_{\theta,B}.
\]
For any two-player Shapley game,
\begin{equation}
\phi_C-\phi_B=v(C)-v(B),
\label{eq:two-player-difference}
\end{equation}
because the joint-coalition term cancels. Therefore
\[
\phi_C^{\cR_0,\star}-\phi_B^{\cR_0,\star}
=
\frac{\gamma_t}{2},
\quad
\phi_B^{\cR_1,\star}-\phi_C^{\cR_1,\star}
=
\frac{\gamma_t}{2}.
\]
For any other access-independent reference with singleton value $c$, the two signed margins are $c-v_0$ and $v_1-c$, whose sum is $\gamma_t$. Their minimum is therefore at most $\gamma_t/2$, with equality only at the midpoint. Hence the optimal symmetric reversal margin for task $t$ is exactly $\gamma_t/2$. Taking the supremum over $(\pi,u)$ and using Step~3 shows that the largest symmetric reversal margin over normalized tasks is $\delta_{0\to1}^{\mathsf A}(\cE)/2$. This proves \cref{thm:valuation-deficiency-duality,cor:resource-separation-reversal,cor:optimal-reversal-margin}. Taking the supremum over finite action alphabets proves \cref{cor:global-duality}. \qed

\subsection{Proof of \cref{cor:universal-train-test-lifting}}
Take a deterministic test input $X=x_0$ and one hypothesis $h_a$ for each decision action $a\in\mathsf A$. Let the test label be the latent variable $\Theta$ and define $\ell_{\mathrm{test}}(h_a,\Theta)=L(\Theta,a)$. A training protocol followed by hypothesis selection is then exactly the original decision procedure, so every subset risk, access advantage, and Shapley value is preserved. \qed

\subsection{Proof of the Shapley shift geometry}
\label{app:proof-shapley-geometry}

Fix arbitrary admissible physical access models $\cR,\cR'$ and a normalized task functional $F_{\theta a}=\pi_\theta u(\theta,a)$. For contributor $i$, Shapley linearity and support-function additivity give
\begin{align*}
\Delta_i^{\cR\to\cR'}(F)
:=&
\phi_i^{\cR',\star}(F)-\phi_i^{\cR,\star}(F)\\
=&
\sum_{S\subseteq N\setminus\{i\}}w_S
\Big[
 h_{\mathfrak P_{\cR'}^{\mathsf A}(\cE_{S\cup\{i\}})}(F)
-h_{\mathfrak P_{\cR'}^{\mathsf A}(\cE_S)}(F)\nonumber\\
&-h_{\mathfrak P_{\cR}^{\mathsf A}(\cE_{S\cup\{i\}})}(F)
+h_{\mathfrak P_{\cR}^{\mathsf A}(\cE_S)}(F)
\Big]\\
=&
h_{\cK_i^+(\cR,\cR')}(F)
-
h_{\cK_i^-(\cR,\cR')}(F).
\end{align*}
In the nested case $\cR_0\subseteq\cR_1$, this is \cref{eq:delta-support-difference}; Shapley efficiency also gives the stated sum over contributors.

For $\xi\in\mathbb R^n$, split $\xi=\xi^+-\xi^-$ coordinatewise. Using nonnegative homogeneity and additivity of support functions,
\begin{align*}
\xi^\top\boldsymbol\Delta_{\cR\to\cR'}(F)
&=
\sum_i \xi_i
\left[h_{\cK_i^+}(F)-h_{\cK_i^-}(F)\right]\\
&=
h_{\cK_\xi^+(\cR,\cR')}(F)
-
h_{\cK_\xi^-(\cR,\cR')}(F).
\label{eq:directional-support-proof}
\end{align*}
For $\xi\neq0$, both directional aggregates have row sum $2\sum_i|\xi_i|$, so their differences lie in the same zero-row-sum conditional-behavior vector space as before; $\xi=0$ is trivial. Because the two aggregates have identical row sums, adding a row-constant functional shifts their two support functions by the same amount and therefore leaves their support difference unchanged.

We use the standard support-function characterization of directed Hausdorff excess in a finite-dimensional normed space. For compact convex sets $\mathcal X,\mathcal Y$ in a common affine space,
\begin{equation}
d_{\rightarrow}(\mathcal X,\mathcal Y)
=
\sup_{\norm{F}_*\le1}
\left[h_{\mathcal X}(F)-h_{\mathcal Y}(F)\right].
\label{eq:directed-support-identity}
\end{equation}
Indeed, $d_{\rightarrow}(\mathcal X,\mathcal Y)$ is the least $r$ such that $\mathcal X\subseteq\mathcal Y+r\mathbb B$; for compact convex sets this inclusion is equivalent to
\[
h_{\mathcal X}(F)
\le
h_{\mathcal Y}(F)+r\norm{F}_*
\quad\text{for all }F.
\]
The quotient dual unit ball on the zero-row-sum conditional-behavior space is the oscillation ball derived in \cref{eq:dual-interface-norm}. Its normalized representatives are \cref{eq:normalized-dual-set}, and Step~3 of the proof of \cref{thm:valuation-deficiency-duality} shows that these representatives are precisely
\[
F_{\theta a}=\pi_\theta u(\theta,a),
\quad
\pi\in\Delta(\Theta),\quad 0\le u\le1.
\]
Applying \cref{eq:directed-support-identity} to the directional support difference above proves
\[
\sup_t \xi^\top\boldsymbol\Delta_{\cR\to\cR'}(t)
=
d_{\rightarrow}(\cK_\xi^+,\cK_\xi^-),
\]
which is \cref{eq:directional-shift-geometry}.

By definition of the closed convex hull,
\[
\sup_{x\in\cV_{\cR\to\cR'}^{\mathsf A}}\xi^\top x
=
\sup_t \xi^\top\boldsymbol\Delta_{\cR\to\cR'}(t).
\]
Combining this identity with \cref{eq:directional-shift-geometry} proves \cref{thm:valuation-shift-body}. \qed

\subsection{Coordinate, ranking, and access-metric corollaries}

For $\xi=e_i$, \cref{eq:directional-shift-geometry} gives \cref{eq:max-appreciation}; for $\xi=-e_i$, the positive and negative parts swap $\cK_i^+$ and $\cK_i^-$ and give \cref{eq:max-devaluation}. Taking the maximum of the two directed simulation gaps gives \cref{eq:max-absolute-shift}. Hausdorff distance vanishes exactly when the two compact closed aggregates coincide, which proves the invariance statement and \cref{cor:native-invariance}. This proves \cref{thm:individual-geometry}.

For $\xi=e_i-e_j$,
\[
\cK_\xi^+=\cK_i^+\oplus\cK_j^-,
\quad
\cK_\xi^-=\cK_i^-\oplus\cK_j^+,
\]
which is \cref{eq:Kij-pm}. Applying the directional theorem to $\xi$ and $-\xi$ proves \cref{eq:max-pair-up,eq:max-pair-down,eq:max-pair-abs}. For a fixed task,
\[
G_{ij}(F)
\ge
-d_{\rightarrow}(\cK_{ij}^-,\cK_{ij}^+),
\]
so adding the weaker-access margin proves \cref{eq:geometric-no-reversal}. This proves \cref{thm:pairwise-geometry}.

Now let $\norm{\cdot}_{\mathsf V}$ be a norm on $\mathbb R^n$. Because norms are symmetric,
\[
\norm{\boldsymbol\phi^{\cR,\star}(t)-\boldsymbol\phi^{\cR',\star}(t)}_{\mathsf V}
=
\norm{\boldsymbol\Delta_{\cR\to\cR'}(t)}_{\mathsf V}.
\]
By duality of finite-dimensional norms and interchange of two suprema,
\begin{align*}
D_{\mathsf V}^{\mathsf A}(\cR,\cR')
&=
\sup_t
\sup_{\norm{\xi}_{\mathsf V,*}\le1}
 \xi^\top\boldsymbol\Delta_{\cR\to\cR'}(t)\\
&=
\sup_{\norm{\xi}_{\mathsf V,*}\le1}
\sup_t
 \xi^\top\boldsymbol\Delta_{\cR\to\cR'}(t)\\
&=
\sup_{\norm{\xi}_{\mathsf V,*}\le1}
 d_{\rightarrow}\!\left(
 \cK_\xi^+(\cR,\cR'),
 \cK_\xi^-(\cR,\cR')
 \right),
\end{align*}
which is \cref{eq:vector-resource-geometry}.

Nonnegativity and symmetry of $D_{\mathsf V}^{\mathsf A}$ are immediate from its definition. For any third access model $\cR''$ and every task $t$, the norm triangle inequality gives
\begin{align}
\norm{\boldsymbol\phi^{\cR,\star}(t)-\boldsymbol\phi^{\cR'',\star}(t)}_{\mathsf V}
\le&
\norm{\boldsymbol\phi^{\cR,\star}(t)-\boldsymbol\phi^{\cR',\star}(t)}_{\mathsf V}\nonumber\\
&+
\norm{\boldsymbol\phi^{\cR',\star}(t)-\boldsymbol\phi^{\cR'',\star}(t)}_{\mathsf V}.
\end{align}
Taking the supremum over tasks proves the pseudometric triangle inequality. The distance vanishes exactly when the full Shapley vectors coincide for every normalized task, so quotienting by that equivalence produces a metric. The scalar functions $|e_i^\top x|$ and $|(e_i-e_j)^\top x|$ obey the same pointwise triangle argument, proving the contributor-specific and pairwise pseudometric statements. This proves \cref{thm:resource-valuation-pseudometric}. \qed

\subsection{Proofs of the deficiency corollaries}

For every normalized task,
\[
0\le g_F(S)\le\delta(S)
\]
by \cref{thm:valuation-deficiency-duality}. Hence
\[
\Delta_i(F)
=
\bbE_{S\sim\Sh_i}[g_F(S\cup\{i\})-g_F(S)]
\le
\bbE_{S\sim\Sh_i}\delta(S\cup\{i\})
\]
and
\[
-\Delta_i(F)
\le
\bbE_{S\sim\Sh_i}\delta(S).
\]
Taking the supremum over tasks and invoking \cref{thm:individual-geometry} gives
\cref{eq:deficiency-appreciation,eq:deficiency-devaluation}. The fixed-task bound
\cref{eq:deficiency-delta} follows before taking the supremum, and the uniform
$\delta_\star^{\mathsf A}$ bound is immediate.

For pairwise ranking drift,
\[
-(\Delta_i-\Delta_j)
\le
\bbE_{S\sim\Sh_i}\delta(S)
+
\bbE_{T\sim\Sh_j}\delta(T\cup\{j\}).
\]
Adding this to the weaker-access margin proves \cref{eq:weighted-no-reversal}. Bounding both expectations by $\delta_\star^{\mathsf A}$ gives the simpler $2\delta_\star^{\mathsf A}$ certificate. \qed

\subsection{Proof of \cref{thm:q-bound}}
Fix $\varepsilon>0$ and choose $\cF=\{\sigma_\theta\}\in\cC_{0,1}^{\mathsf A}$ such that
\[
\sup_\theta T(\rho_\theta,\sigma_\theta)
\le
q_{0,1}^{\mathsf A}(\cE)+\varepsilon.
\]
The function $P\mapsto\inf_{Q\in\mathfrak P_{\cR_0}^{\mathsf A}(\cE)}\norm{P-Q}_{\infty,\TV}$ is continuous, so taking the supremum over the closed accessible-output set is equivalent to taking the supremum over its dense set of actually implementable convex-randomized output behaviors. It therefore suffices to fix an implementable richer-access protocol $M$ on $\cE$. Let $P_{\rho}$ be its output behavior on $\cE$ and $P_{\sigma}$ the output behavior obtained by applying the same protocol to $\cF$. Because $\delta_{0\to1}^{\mathsf A}(\cF)=0$, for every $\eta>0$ there exists a weaker-access protocol with output behavior $Q_\sigma$ on $\cF$ satisfying
\[
\sup_\theta\TV(P_{\theta,\sigma},Q_{\theta,\sigma})\le\eta.
\]
Apply that same weak protocol to $\cE$ and call its output behavior $Q_\rho$. For every $\theta$,
\begin{align*}
\TV(P_{\theta,\rho},Q_{\theta,\rho})
&\le
\TV(P_{\theta,\rho},P_{\theta,\sigma})
+
\TV(P_{\theta,\sigma},Q_{\theta,\sigma})\\
&\quad+
\TV(Q_{\theta,\sigma},Q_{\theta,\rho})\\
&\le
T(\rho_\theta,\sigma_\theta)+\eta+T(\rho_\theta,\sigma_\theta),
\end{align*}
where the last inequality is contractivity of trace distance under measurement. Taking the supremum over $\theta$, infimum over weak-access output behaviors, and supremum over strong-access output behaviors gives
\[
\delta_{0\to1}^{\mathsf A}(\cE)
\le
2(q_{0,1}^{\mathsf A}(\cE)+\varepsilon)+\eta.
\]
Let $\varepsilon,\eta\downarrow0$. The individual Shapley and ranking bounds follow from \cref{thm:deficiency-premium,cor:no-reversal}. \qed

\section{One-Copy Construction and Proofs}
\label{app:one-copy-construction}

This appendix records the calculations used in \cref{sec:reversal}.

\subsection{From class identification to supervised test risk}

Take $X\sim\operatorname{Unif}\{1,2,3\}$ with 
\[
h_0=(0,0,0),\quad
h_1=(0,1,1),\quad
h_2=(1,0,1),
\]
and let the target be $Y=h_\Theta(X)$. For any $j\neq r$,
\begin{equation}
\Pr_X[h_j(X)\neq h_r(X)]=\frac23.
\label{eq:equidistant-code}
\end{equation}
Condition on any measurement outcome $Z=z$ from the contributor data and write
\[
\pi_j(z)=\bbP(\Theta=j\mid Z=z).
\]
If the learner outputs $h_r$, its conditional test risk is
\[
\sum_j\pi_j(z)\Pr_X[h_j(X)\neq h_r(X)]
=
\frac23[1-\pi_r(z)].
\]
The Bayes-optimal choice is therefore a posterior maximizer, and averaging over $Z$ gives
\[
R_{\cR}^{\star}(S)
=
\frac23\left[
1-\bbE_Z\max_r\pi_r(Z)
\right].
\]
The quantity $\bbE_Z\max_r\pi_r(Z)$ is exactly the optimal success probability $p_{\cR}(S)$ for identifying $\Theta$, proving \cref{eq:risk-success}. With no data, $p(\varnothing)=1/3$, hence $R_{\emptyset}=4/9$ and \cref{eq:value-success} follows.

\subsection{Contributor states and success probabilities}

Contributor $A$ is the classical erasure register
\[
\rho_{\theta,A}
=
s\,\ketbra{\theta}{\theta}
+
(1-s)\ketbra{e}{e},
\quad
s=\frac{37}{40}.
\]
It reveals $\Theta$ with probability $s$ and otherwise leaves the posterior uniform. Therefore
\[
p_A=s+\frac{1-s}{3}=\frac{19}{20}.
\]

For contributor $B$,
\[
\ket{s_0}=\ket0,\quad
\ket{s_1}=-\tfrac12\ket0+\tfrac{\sqrt3}{2}\ket1,\quad
\ket{s_2}=-\tfrac12\ket0-\tfrac{\sqrt3}{2}\ket1,
\]
and
\[
\ket{D_j}=\ket{s_j}\otimes\ket{s_j}.
\]
The known optimal discrimination probabilities for the double-trine ensemble are
\[
p_B^{\LOCC}
=
\frac12+\frac{\sqrt3}{4},
\quad
p_B^G
=
\frac12+\frac{\sqrt2}{3}.
\]
These satisfy
\[
p_B^{\LOCC}<\frac{19}{20}<p_B^G.
\]

For the joint contributor state
\[
\rho_{\theta,AB}
=
\rho_{\theta,A}\otimes\ketbra{D_\theta}{D_\theta},
\]
the orthogonal classical register of $A$ can be read first without disturbing $B$. With probability $s$ it reveals $\Theta$ exactly; with probability $1-s$ it returns the erasure symbol, after which the optimal $B$-measurement is used. Hence
\[
p_{\cR}(AB)=s+(1-s)p_B^{\cR}.
\]
For the two access models,
\[
p_{\LOCC}(AB)
=
\frac{77}{80}+\frac{3\sqrt3}{160},
\quad
p_G(AB)
=
\frac{77}{80}+\frac{\sqrt2}{40}.
\]

\subsection{Coalition values and Shapley values}

Using
\[
v_{\cR}^{\star}(S)
=
\frac23\left[p_{\cR}(S)-\frac13\right]
\]
gives
\[
v(\varnothing)=0,
\quad
v(A)=\frac{37}{90},
\]
and
\[
v_{\LOCC}(B)
=
\frac19+\frac{\sqrt3}{6},
\quad
v_G(B)
=
\frac{1+2\sqrt2}{9},
\]
together with
\[
v_{\LOCC}(AB)
=
\frac{151}{360}+\frac{\sqrt3}{80},
\quad
v_G(AB)
=
\frac{151}{360}+\frac{\sqrt2}{60}.
\]

For a two-player game,
\[
\phi_A
=
\frac12v(A)+\frac12[v(AB)-v(B)],
\quad
\phi_B
=
\frac12v(B)+\frac12[v(AB)-v(A)].
\]
Therefore
\[
\phi_A^{\LOCC}
=
\frac{259}{720}-\frac{37\sqrt3}{480},
\quad
\phi_B^{\LOCC}
=
\frac{43}{720}+\frac{43\sqrt3}{480},
\]
while
\[
\phi_A^G
=
\frac{259}{720}-\frac{37\sqrt2}{360},
\quad
\phi_B^G
=
\frac{43}{720}+\frac{43\sqrt2}{360}.
\]
Numerically,
\[
(\phi_A^{\LOCC},\phi_B^{\LOCC})
\approx
(0.2262,0.2149),
\quad
(\phi_A^G,\phi_B^G)
\approx
(0.2144,0.2286).
\]
Thus the top-ranked contributor changes from $A$ under one-way LOCC to $B$ under global access.

As a check, subtracting the two-player formulas gives
\[
\phi_A-\phi_B=v(A)-v(B),
\]
so the sign change is already visible from the singleton ordering, while the joint-coalition calculation above is what fixes the individual Shapley values themselves.

\section{Proofs for the Information-Theoretic Decomposition}

\subsection{Proof of \cref{prop:info-shapley}}

The chain rule for quantum mutual information gives
\[
I(Y:Q_{S\cup\{i\}})
-
I(Y:Q_S)
=
I(Y:Q_i\mid Q_S).
\]
Substituting this marginal into the Shapley formula yields \cref{eq:cmi-shapley}. Strong subadditivity implies nonnegativity of quantum conditional mutual information. Shapley efficiency gives
\[
\sum_i\psi_i^\chi
=
v_\chi(N)-v_\chi(\varnothing)
=
I(Y:Q_N).
\]
Finally,
\[
A_{\cR}=v_\chi-D_{\cR}
\]
and Shapley linearity give \cref{eq:intrinsic-operational}. \qed

\section{Proofs for Fixed-Learner Finite-Copy Valuation}

\subsection{Proof of \cref{thm:normative-operational-bridge}}
\label{app:proof-normative-operational-bridge}
Define the coalition suboptimality
\[
e(S)=v_{\cR}^{\star}(S)-v_{\cA}(S).
\]
By assumption, $e(S)\in[0,\eta]$ for every coalition. Shapley linearity gives
\[
\phi_i^{\cR,\star}-\phi_i^{\cA}
=
\Sh_i(e)
=
\sum_{S\subseteq N\setminus\{i\}}
w_S[e(S\cup\{i\})-e(S)].
\]
Since both terms in each bracket lie in $[0,\eta]$,
\[
|e(S\cup\{i\})-e(S)|\le\eta.
\]
The Shapley weights are nonnegative and sum to one, hence
\[
|\phi_i^{\cR,\star}-\phi_i^{\cA}|\le\eta.
\]
The estimator bound follows from the triangle inequality. \qed

\subsection{Proof of \cref{cor:finite-copy-resource-certificate}}
On the simultaneous event in the statement, each estimated pairwise gap differs from the corresponding best-achievable gap by at most $2\tau$. Therefore the true symmetric reversal margin $m$ satisfies
\[
m\ge[\widehat m-2\tau]_+.
\]
For the fixed task, \cref{cor:optimal-reversal-margin} and \cref{thm:valuation-deficiency-duality} imply
\[
m\le\frac12\delta_{0\to1}^{\mathsf A}(\cE_B).
\]
Combining gives \cref{eq:finite-copy-deficiency-certificate}. Finally \cref{thm:q-bound} gives $\delta_{0\to1}^{\mathsf A}\le2q_{0,1}^{\mathsf A}$, hence \cref{eq:finite-copy-resource-insensitivity-certificate}. \qed

\subsection{Proof of \cref{thm:valuation-observable}}

Fix coalition $S$. The entire pipeline
\begin{align}
&\varrho_S
\longmapsto
\text{training protocol }\cA_S
\longmapsto
\text{hypothesis}\nonumber\\
&\longmapsto
\text{independent test evaluation}\nonumber
\end{align}
is a fixed quantum-to-classical experiment followed by a score in $[0,1]$. By linearity of quantum channels and expectation, the expected test score is an affine linear functional of $\varrho_S$. The Riesz representation on finite-dimensional Hermitian operators therefore gives an effect $0\le F_S\le\1_S$ whose expectation equals that score:
\[
\text{expected score from }S=\Tr(F_S\varrho_S).
\]
Since the omitted subsystems are normalized,
\[
\Tr(F_S\varrho_S)
=
\Tr((F_S\otimes\1_{S^c})\varrho_D)
=
\Tr(\widetilde F_S\varrho_D).
\]
The no-data baseline cancels in every Shapley marginal, so inserting these score differences into the Shapley formula gives
\begin{align*}
\phi_i^{\cA}
&=
\sum_{S\subseteq N\setminus\{i\}}
w_S
\Tr[
(\widetilde F_{S\cup\{i\}}-\widetilde F_S)\varrho_D]\\
&=
\Tr(\Omega_i\varrho_D).
\end{align*}
For two effects $A,B$, $-\1\le A-B\le\1$, hence $\norm{A-B}_\infty\le1$. The Shapley weights are nonnegative and sum to one, so
\[
\norm{\Omega_i}_\infty
\le
\sum_{S\subseteq N\setminus\{i\}}w_S
\norm{\widetilde F_{S\cup\{i\}}-\widetilde F_S}_\infty
\le1.
\]
\qed

\subsection{Proof of \cref{prop:single-copy-complexity}}

For the upper bound, spectrally measure $\Omega_i$. Every outcome lies in an interval of length $W_i$, and its expectation is $\phi_i$. Hoeffding's inequality gives
\[
\bbP(|\widehat\phi_i-\phi_i|>\epsilon)
\le
2\exp\!\left(-\frac{2K\epsilon^2}{W_i^2}\right).
\]
Solving for $K$ proves the upper bound.

For the lower bound, fix any $0<W\le1$ and set
\[
E_W=W\ketbra{0}{0},
\quad
\rho_\pm
=
\diag\!\left(
\frac12\pm\frac{2\epsilon}{W},
\frac12\mp\frac{2\epsilon}{W}
\right).
\]
The assumption $\epsilon\le W/8$ makes both states valid and keeps their Bernoulli parameters in $[1/4,3/4]$. The observable $E_W$ has spectral width exactly $W$, and
\[
\Tr(E_W\rho_+)-\Tr(E_W\rho_-)=4\epsilon.
\]
Thus any estimator that is $\epsilon$-accurate on both states with failure probability at most $p_{\mathrm{fail}}$ yields, by midpoint thresholding, a binary test with both error probabilities at most $p_{\mathrm{fail}}$.

Because $\rho_+$ and $\rho_-$ commute, measuring the common eigenbasis on every replica produces a classical Bernoulli sequence that is sufficient for every collective quantum measurement: any POVM on the commuting product states is a stochastic post-processing of that sequence. Put $a=2\epsilon/W\le1/4$. The one-sample symmetrized KL divergence is
\[
4a\log\frac{1+2a}{1-2a}
\le32a^2,
\]
where the inequality uses $\log((1+x)/(1-x))\le4x$ for $0\le x\le1/2$. On the other hand, the binary test with both errors at most $p_{\mathrm{fail}}\le1/4$ implies, by data processing to its decision bit,
\[
D_{\mathrm{KL}}^{\mathrm{sym}}
\ge
2(1-2p_{\mathrm{fail}})\log\frac{1-p_{\mathrm{fail}}}{p_{\mathrm{fail}}}.
\]
KL additivity over the sufficient Bernoulli sequence therefore gives
\[
K
=\Omega\!\left(
\frac{W^2}{\epsilon^2}\log\frac{1}{p_{\mathrm{fail}}}
\right).
\]
This is a valid one-player \raqdv\ instance: take $u(\varnothing)=0$ and realize $u(\{1\})=\Tr(E_W\rho)$ by the two-outcome effect $E_W$, assigning unit score to that outcome and zero otherwise. The utility lies in $[0,1]$ because $0<W\le1$, and the Shapley observable is exactly $\Omega_1=E_W$. \qed

\subsection{Proof of \cref{thm:shadow-reduction}}

By \cref{thm:valuation-observable}, each fixed-learner valuation is exactly the expectation of the known Hermitian observable $\Omega_i$. Therefore any many-observable estimation guarantee applies verbatim to the family $\{\Omega_i\}$. The stated classical-shadow scaling is the standard guarantee in terms of the maximum shadow norm of the target family \citep{huang2020shadows}. \qed

\subsection{Proof of \cref{prop:degree-classical}}

Fix $m$ and suppose a protocol in $\cR^{(k)}[\mathsf B(m)]$ is applied to the classically readable training states of $D_m$. Measure every supplied copy independently in the common readable basis and retain the complete classical outcome record. For states diagonal in that product basis, this record is a sufficient classical statistic for every subsequent coherent block measurement: the outcome distribution of any allowed $k$-copy POVM is a stochastic post-processing of the same record, by the argument of \cref{thm:classical-collapse}. The simulation uses the same number of state preparations and only degree-one quantum measurements. Therefore
\[
v_{\cR^{(1)}}^{\star,\mathsf B}(D_m)
\ge
v_{\cR^{(k)}}^{\star,\mathsf B}(D_m)
\]
for every $m$ and $k$ (the reverse inequality follows from access-model nesting). If some finite degree reaches $\tau$ for all sufficiently large $m$, degree one reaches the same threshold for those $m$, proving $d_{\tau,\mathsf B}(\mathbf D)=1$. \qed

\subsection{Proof of \cref{cor:degree-hierarchy}}

Let $u_\emptyset$ be the common no-data optimum in the bounded utility encoding assumed by the corollary. Choose a polynomial budget $\mathsf B(m)$ large enough to contain the efficient $k$-copy protocol and choose a positive threshold
\[
\tau\in
\bigl(u_{\rm lo}-u_\emptyset,\;u_{\rm hi}-u_\emptyset\bigr),
\]
which is nonempty because $u_{\rm hi}>u_{\rm lo}$ and $u_\emptyset\le u_{\rm lo}$. For all sufficiently large $m$,
\[
v_{\cR^{(k)}}^{\star,\mathsf B}(D_m)
=U_{\cR^{(k)}}^{\star,\mathsf B}(D_m)-u_\emptyset
\ge u_{\rm hi}-u_\emptyset>\tau.
\]
For every $j<k$, the lower side of the constant-gap hierarchy gives
\[
v_{\cR^{(j)}}^{\star,\mathsf B}(D_m)
=U_{\cR^{(j)}}^{\star,\mathsf B}(D_m)-u_\emptyset
\le u_{\rm lo}-u_\emptyset<\tau.
\]
Therefore the smallest coherent copy degree that eventually reaches the target value within the polynomial budget is exactly $k$, i.e. $d_{\tau,\mathsf B}(\mathbf D)=k$. No new copy-complexity separation is used. \qed

\section{Proofs for Valuation--Disturbance}

\subsection{Proof of \cref{prop:no-universal-task}}

Let
\[
\rho_\theta=\tau_\theta^V\otimes\sigma_\theta^T.
\]
Choose a nonconstant valuation that depends only on $\tau_\theta^V$ and a valuation instrument of the form
\[
\mathcal I_{\mathrm{val}}=\mathcal I_V\otimes\id_T.
\]
The protocol may extract nontrivial valuation information and arbitrarily disturb $V$, but the reduced state on $T$ remains exactly $\sigma_\theta^T$. Any fixed future learning/decision task whose optimal utility depends only on $T$ has identical pre- and post-valuation performance. Therefore no strictly positive universal task-independent damage lower bound exists. \qed

\subsection{Proof of \cref{lem:valuation-distinguishability}}

Let
\[
m=\frac{\phi(\rho_0)+\phi(\rho_1)}2.
\]
Because the true values differ by more than $2\epsilon$, the two intervals
\[
[\phi(\rho_0)-\epsilon,\phi(\rho_0)+\epsilon]
\quad\text{and}\quad
[\phi(\rho_1)-\epsilon,\phi(\rho_1)+\epsilon]
\]
lie on opposite sides of $m$. Thresholding the estimator at $m$ therefore distinguishes the two input states with error at most $p_{\mathrm{fail}}$ under either hypothesis. The optimal classical binary-testing error for equal priors is
\[
P_e^\star=\frac12(1-\TV(P_{Z|0},P_{Z|1})).
\]
Since a test with error at most $p_{\mathrm{fail}}$ exists,
\[
\TV(P_{Z|0},P_{Z|1})\ge1-2p_{\mathrm{fail}}.
\]
\qed

\subsection{Proof of \cref{thm:residual-no-free}}

Take a Stinespring dilation of the residual channel $\cN$ whose complementary output includes the valuation transcript $Z$. By data processing of trace distance, the complementary outputs on $\rho_0$ and $\rho_1$ have trace distance at least that visible in $Z$:
\[
\frac12
\norm{
\cN^c(\rho_0)-\cN^c(\rho_1)
}_1
\ge
\TV(P_{Z|0},P_{Z|1})
\ge1-2p_{\mathrm{fail}}.
\]
For any constant channel $\cS$,
\begin{align*}
\norm{\cN^c-\cS}_{\diamond}
&\ge
\frac12
\norm{
\cN^c(\rho_0)-\cN^c(\rho_1)
}_1\\
&\ge
1-2p_{\mathrm{fail}},
\end{align*}
where the first inequality follows from the triangle inequality applied to the two inputs. The information--disturbance theorem of \citet{ksw2006} gives, under the same unnormalized norm convention,
\[
\inf_{\cS\ \mathrm{constant}}
\norm{\cN^c-\cS}_{\diamond}
\le
2\sqrt{
\inf_{\cR_{\mathrm{rec}}}
\norm{\cR_{\mathrm{rec}}\circ\cN-\id_Q}_{\diamond}
}.
\]
Combining the two inequalities yields
\[
\eta_{\mathrm{rec}}(\cN)
\ge
\frac{(1-2p_{\mathrm{fail}})^2}{4}.
\]
A nonzero recovery deficiency means the residual experiment is not statistically equivalent to the identity experiment. The ancilla-assisted quantum randomization criteria therefore guarantee the existence of a decision problem in the corresponding comparison class with a strict operational gap \citep{jencova2016comparison}. We deliberately do not assign the same numerical constant to an arbitrary predetermined supervised loss; such a claim would require specifying that loss inside the same decision class. \qed

\subsection{Proof of \cref{thm:gentle-lower}}

Let $\ket{+}$ and $\ket{-}$ be eigenvectors of $\Omega$ for $\lambda_{\max}$ and $\lambda_{\min}$. Set
\[
s=\frac{2\epsilon}{W_\Omega},
\]
and define
\[
\rho_+
=
\left(\frac12+s\right)\ketbra{+}{+}
+
\left(\frac12-s\right)\ketbra{-}{-},
\]
\[
\rho_-
=
\left(\frac12-s\right)\ketbra{+}{+}
+
\left(\frac12+s\right)\ketbra{-}{-}.
\]
The assumption $\epsilon\le W_\Omega/8$ gives $s\le1/4$, so both are valid states. Their valuation difference is
\[
\phi_\Omega(\rho_+)-\phi_\Omega(\rho_-)
=
2sW_\Omega
=
4\epsilon.
\]
Their trace distance is
\[
T(\rho_+,\rho_-)
=
\frac12\norm{\rho_+-\rho_-}_1
=
2s
=
\frac{4\epsilon}{W_\Omega}.
\]

Let $P_+$ and $P_-$ denote the complete classical transcript laws of the $K$-copy valuation protocol on $\rho_+^{\otimes K}$ and $\rho_-^{\otimes K}$. Thresholding an $\epsilon$-accurate estimator at the midpoint of the two true valuations produces a binary test. If $\mathsf E_+$ denotes the event that this test chooses the $+$ state, then
\[
p:=P_+(\mathsf E_+)\ge1-p_{\mathrm{fail}},
\quad
q:=P_-(\mathsf E_+)\le p_{\mathrm{fail}}.
\]
By data processing of KL divergence under the binary map $\mathbf 1_{\mathsf E_+}$,
\begin{align}
D_{\mathrm{KL}}^{\mathrm{sym}}(P_+,P_-)
&\ge
D_{\mathrm{KL}}^{\mathrm{sym}}
\big(
\mathrm{Ber}(p),\mathrm{Ber}(q)
\big)\nonumber\\
&=
(p-q)
\log
\frac{p(1-q)}{q(1-p)}\nonumber\\
&\ge
2(1-2p_{\mathrm{fail}})
\log\frac{1-p_{\mathrm{fail}}}{p_{\mathrm{fail}}}.
\label{eq:binary-kl-confidence}
\end{align}
The last inequality uses $p-q\ge1-2p_{\mathrm{fail}}$ and
\[
\frac{p(1-q)}{q(1-p)}
\ge
\left(\frac{1-p_{\mathrm{fail}}}{p_{\mathrm{fail}}}\right)^2.
\]

For the non-adaptive product measurement in the theorem, the transcript law is a product law and symmetrized KL divergence adds over copies. Applying \cref{eq:gentle-qdpi} to each copy gives
\begin{align*}
D_{\mathrm{KL}}^{\mathrm{sym}}(P_+,P_-)
&\le
K
\left(
\frac{8\alpha}{(1-2\alpha)^2}
\right)^2
T(\rho_+,\rho_-)^2\\
&=
\frac{
1024\,K\alpha^2\epsilon^2
}{
W_\Omega^2(1-2\alpha)^4
}.
\end{align*}
Combining this upper bound with \cref{eq:binary-kl-confidence} yields
\[
K
\ge
\frac{
W_\Omega^2(1-2\alpha)^4
}{
512\,\alpha^2\epsilon^2
}
(1-2p_{\mathrm{fail}})
\log\frac{1-p_{\mathrm{fail}}}{p_{\mathrm{fail}}},
\]
which is \cref{eq:gentle-lower-bound}. For $p_{\mathrm{fail}}\le1/4$,
\[
(1-2p_{\mathrm{fail}})\log\frac{1-p_{\mathrm{fail}}}{p_{\mathrm{fail}}}
=
\Theta\!\left(\log\frac{1}{p_{\mathrm{fail}}}\right),
\]
and for $\alpha\le\alpha_0<1/2$, the factor $(1-2\alpha)^4$ is bounded below by a positive constant depending only on $\alpha_0$. This proves \cref{eq:gentle-confidence-lower}. \qed

\subsection{Proof of \cref{cor:gentle-upper}}

Let
\[
E=
\frac{\Omega-\lambda_{\min}\1}{W_\Omega},
\quad
0\le E\le\1,
\]
and write
\[
p=\Tr(E\rho)
=
\frac{\phi_\Omega(\rho)-\lambda_{\min}}{W_\Omega}.
\]
Choose a standard Naimark dilation of the binary POVM $\{E,\1-E\}$. Thus there are an ancilla initialized in $\ket{0}$, a unitary $V_{\mathrm N}$ on system plus ancilla, and a binary projector $\Pi$ on the enlarged space such that, for every input state $\rho$,
\[
\Tr\!\left[\Pi\,V_{\mathrm N}(\rho\otimes\ketbra{0}{0})V_{\mathrm N}^\dagger\right]
=
\Tr(E\rho)=p.
\]
Apply the quantum Label Switch mechanism of \citet{butucea2025gentle} to the projective measurement $\{\Pi,\1-\Pi\}$, taking its state class to be the full enlarged state space. If $\widetilde\rho=V_{\mathrm N}(\rho\otimes\ketbra{0}{0})V_{\mathrm N}^\dagger$ and $\widetilde\rho_y'$ is the post-qLS enlarged state conditioned on transcript $Y=y$, their gentleness guarantee gives
\[
T(\widetilde\rho_y',\widetilde\rho)\le\alpha.
\]
Decode by $V_{\mathrm N}^\dagger$ and discard the ancilla. Writing
\[
\rho_y'=\Tr_{\mathrm{anc}}\!\left[V_{\mathrm N}^\dagger\widetilde\rho_y'V_{\mathrm N}\right],
\]
unitary invariance and CPTP contractivity of trace distance yield
\[
T(\rho_y',\rho)\le\alpha.
\]
Hence the induced single-copy instrument is $\alpha$-gentle on the full original state space $\cS(\cH)$, exactly as required by \cref{def:gentle-val-complexity}. The Naimark dilation preserves the underlying binary probability $p$, so the qLS transcript $Y$ has the same known affine mean
\[
\bbE_\rho Y
=
\frac{1-\beta_\alpha}{2}
+
\beta_\alpha p,
\quad
\beta_\alpha
=
\frac{2\alpha}{1+\alpha^2}.
\]
Hence
\[
\widehat p
=
\frac{
\overline Y-(1-\beta_\alpha)/2
}{
\beta_\alpha
}
\]
is unbiased. Since $Y\in[0,1]$, Hoeffding's inequality gives
\[
\bbP\!\left(
|\widehat p-p|>\frac{\epsilon}{W_\Omega}
\right)
\le
2\exp\!\left(
-\frac{
2K\beta_\alpha^2\epsilon^2
}{
W_\Omega^2
}
\right).
\]
Thus \cref{eq:gentle-upper-bound} suffices, and multiplying $\widehat p$ by $W_\Omega$ and adding $\lambda_{\min}$ gives the corresponding valuation estimator. Finally, for $0<\alpha\le\alpha_0<1/2$,
\[
\beta_\alpha
=
\Theta_{\alpha_0}(\alpha),
\]
so the upper bound together with \cref{thm:gentle-lower} yields the confidence-tight minimax law \cref{eq:gentle-minimax-tight}. \qed

\makeatletter
\let\@bibitemShut\@empty
\makeatother

\bibliography{refs}

\end{document}